\documentclass[%
 reprint,
 amsmath,amssymb,
 aps,
prb,
 superscriptaddress
]{revtex4-2}

\usepackage{graphicx}
\usepackage{dcolumn}
\usepackage{bm}
\usepackage{color}
\newcommand{\1}{\mbox{1}\hspace{-0.25em}\mbox{l}}

\begin{document}

\preprint{APS/123-QED}

\title{Non-Bloch band theory of generalized eigenvalue problems}

\author{Kazuki Yokomizo}
\affiliation{Department of Physics, The University of Tokyo, 7-3-1 Hongo, Bunkyo-ku, Tokyo, 113-0033, Japan}
\author{Taiki Yoda}
\affiliation{Department of Physics, Tokyo Institute of Technology, 2-12-1 Ookayama, Meguro-ku, Tokyo, 152-8551, Japan}
\author{Yuto Ashida}%
\affiliation{Department of Physics, The University of Tokyo, 7-3-1 Hongo, Bunkyo-ku, Tokyo, 113-0033, Japan}
\affiliation{Institute for Physics of Intelligence, The University of Tokyo, 7-3-1 Hongo, Tokyo 113-0033, Japan}




%
\begin{abstract}
Waves in a variety of fields in physics, such as mechanics, optics, spintronics, and nonlinear systems, obey generalized eigenvalue equations. To study non-Hermitian physics of those systems, in this paper, we construct a non-Bloch band theory of generalized eigenvalue problems. Specifically, we show that eigenvalues of a transfer matrix lead to a certain condition imposed on the generalized Brillouin zone, which allows us to develop a theory to calculate the continuum bands. As a concrete example, we examine the non-Hermitian skin effect of photonic crystals composed of chiral metamaterials by invoking our theoretical framework. When the medium has circularly polarized eigenmodes, we find that each eigenmode localizes at either of the edges, depending on whether it is left- or right-circularly polarized. In contrast, when the medium only has linearly polarized eigenmodes, every eigenmode localizes to the edge of the same side independent of its polarization. We demonstrate that the localization lengths of those eigenmodes can be determined from the chiral parameters and eigenfrequencies of the photonic crystal.
\end{abstract}
\pacs{Valid PACS appear here}
\maketitle
%
%

\section{\label{sec1}Introduction}
Non-Hermitian physics has been gaining growing interest over the past few years, in which one can find rich phenomena with no counterparts in Hermitian regimes~\cite{Ashida2020,Bergholtz2021}. Non-Hermitian systems belong to a class of nonequilibrium systems which can be described by certain non-Hermitian operators, whose eigenvalues characterize the dynamical behaviors. A prominent example is the non-Hermitian skin effect, in which numerous eigenstates are localized at boundaries of non-Hermitian systems~\cite{Martinez2018,Kunst2018,Yao2018,Song2019,Borgnia2020,Okuma2020,Zhang2020,Yi2020,Brandenbourger2019,Xiao2020,Weidemann2020,Helbig2020,Hofmann2020,Ghatak2020,Chen2021,Zhang2021,Wang2022,Liang2022,Gu2022}. Accordingly, it must be necessary to modify the conventional Bloch band theory in such a way that the continuum energy bands reproduce the energy levels of non-Hermitian systems with open boundary conditions. The resulting theory is called the non-Bloch band theory, which has been successful to calculate the continuum energy bands in the presence of the non-Hermitian skin effect~\cite{Yao2018,Yokomizo2019,Yokomizo2020,Kawabata2020,Yang2020,Yokomizo2022,Hu2023}.

On another front, there exist several physical platforms that can be described by generalized eigenvalue equations in, e.g., mechanics~\cite{Inoue2019}, optics~\cite{Haldane2008,Raman2010}, spintronics~\cite{Shidou2013,Shidou2013v2}, and nonlinear systems~\cite{Isobe2023v2}. Notably, even if all operators appearing in generalized eigenvalue equations are Hermitian, the corresponding eigenvalues can be complex-valued~\cite{Isobe2021,Isobe2023}. Indeed, it has been proposed that bosonic Bogoliubov--de Gennes systems exhibit the non-Hermitian skin effect in the regime where the pseudo-Hermiticity is broken~\cite{McDonald2018,Yokomizo2021}. Aside from this example, several non-Hermitian phenomena in various systems described by generalized eigenvalue equations have been investigated~\cite{Rosa2020,Braghini2021,Yan2021}. It remains, however, unclear how and even whether the non-Bloch band theory can be extended to generalized eigenvalue problems.

In our paper, we develop a non-Bloch band theory to study non-Hermitian waves in a one-dimensional (1D) periodic medium described by a generalized eigenvalue equation. In the context of the non-Bloch band theory, the continuum bands can be obtained by determining the generalized Brillouin zone for the complex-valued Bloch wave number. We show that the eigenvalues of a transfer matrix of the medium lead to a certain condition imposed on the generalized Brillouin zone. The generalized theory proposed here allows us to explore the non-Hermitian skin effect in photonic crystals composed of chiral metamaterials. While parity-time-symmetric metamaterials have been investigated in, e.g., Refs.~\cite{Gear2015,Droulias2019}, our motivation is to go beyond the previous works. We confirm that the continuum bands obtained from the generalized Brillouin zone reproduce the eigenvalues of finite-size systems. In particular, when the medium has circularly polarized eigenmodes, we find that each eigenmode can localize at either of the edges, depending on whether it is left- or right-circularly polarized. This finding suggests the interesting possibility of controlling which side excitation modes localize at simply by changing a polarization of incident light while keeping the medium unchanged. In contrast, when the medium only has linearly polarized eigenmodes, our analysis shows that every eigenmode localizes at one edge of the system.

Before concluding this section, we shall compare the present paper with previous studies \cite{Yokomizo2022,Yoda2023}. The latter focused on dielectric media with optical loss and investigated electromagnetic waves in the two-dimensional plane where the dielectric tensor is anisotropic. There, one of the key findings was that the combination of the anisotropy and the loss gives rise to the non-Hermitian skin effect, and the resulting localization depends on wave numbers of a specified direction. In photonic crystals composed of the dielectric media, for instance, the non-Bloch band theory has revealed that all the continuum bands share the common generalized Brillouin zone. Meanwhile, in the present paper, we investigate standing waves of electromagnetic waves forming along the stacking direction of photonic crystals composed of chiral metamaterials. Here, the key ingredient of the occurrence of the non-Hermitian skin effect is chirality. Indeed, the localization lengths of the skin modes depend on chirality parameters and eigenfrequencies of the photonic crystals. In addition, it is remarkable that the multiple generalized Brillouin zones appear, depending on the continuum bands, in contrast to the previous studies.

The rest of this paper is organized as follows. We introduce a generalized eigenvalue equation, explain the construction of a transfer matrix, and show the condition for the generalized Brillouin zone in Sec.~\ref{sec2}. We next study the non-Hermitian skin effect of linearly and circularly polarized eigenmodes of the photonic crystals by invoking the non-Bloch band theory in Sec.~\ref{sec3}. Finally, in Sec.~\ref{sec4}, we summarize the results and comment on the perspective of this paper.

%
%

\section{\label{sec2}Non-Bloch band theory}
We develop a theory to calculate the continuum bands in 1D continuous periodic models described by a generalized eigenvalue equation. Our framework is based on a transfer matrix method, where the eigenvalues of a transfer matrix lead to the condition for the generalized Brillouin zone.

%
%

\subsection{\label{sec2-1}Generalized eigenvalue equation}
We study the waves in a 1D spatially periodic medium which can be modeled by a generalized eigenvalue equation. We denote a lattice constant by $a$ and take the convention of the time dependence of the wavefunction to be $e^{-i\omega t}$. Furthermore, we assume that the wave function is written as a multicomponent vector, $\Psi\left(z\right)=\left(\psi_1\left(z\right),\dots,\psi_{2n}\left(z\right)\right)^{\rm T}$, with $n\in{\mathbb N}$ and $z\in{\mathbb R}$. The generalized eigenvalue equation governing our setup then reads
\begin{equation}
\frac{d}{dz}\Psi\left(z\right)=i\frac{\omega}{c}A\left(z\right)\Psi\left(z\right),
\label{eq1}
\end{equation}
where $c$ is a positive constant, and $A\left(z\right)$ is a $2n\times2n$ matrix satisfying $A\left(z+a\right)=A\left(z\right)$. The form of Eq.~(\ref{eq1}) naturally appears by rewriting the Maxwell's equations, as demonstrated later.

We aim to investigate a method to calculate the continuum bands, including the asymptotic eigenvalues of the system with open boundary conditions in the limit of a large system size. To this end, we focus on the plane-wave expansion of the wavefunction given by
\begin{equation}
\Psi\left(z\right)=\sum_n\tilde{\Psi}\left(k+\frac{2n\pi}{a}\right)\exp\left[i\left(k+\frac{2n\pi}{a}\right)z\right],
\label{eq2}
\end{equation}
where $k$ represents the Bloch wavenumber. The generalized eigenvalue equation can then be rewritten in the form of a secular equation as follows:
\begin{eqnarray}
&&\left(k+\frac{2n\pi}{a}\right)\tilde{\Psi}\left(k+\frac{2n\pi}{a}\right) \nonumber\\
&&-\omega\sum_{n^\prime}\tilde{A}_{n-n^\prime}\tilde{\Psi}\left(k+\frac{2n^\prime\pi}{a}\right)=0,
\label{eq3}
\end{eqnarray}
where $\tilde{A}_n$ is a Fourier coefficient of $A\left(z\right)$. The localized nature due to the non-Hermitian skin effect can be taken into account by considering a complex-valued Bloch wavenumber $k$~\cite{Yao2018,Yokomizo2019}. Therefore, it is necessary to determine the generalized Brillouin zone to obtain the continuum bands by solving Eq.~(\ref{eq3}).

%
%

\subsection{\label{sec2-2}Transfer matrix}
We next utilize a transfer matrix to develop the non-Bloch band theory which is applicable to the system described by Eq.~(\ref{eq1}). Specifically, we can define the transfer matrix as follows:
\begin{equation}
\Psi\left(z+a\right)=T\Psi\left(z\right),
\label{eq4}
\end{equation}
where the wave function in a unit cell is transferred to that in the next unit cell by the transfer matrix $T$, i.e., the transfer matrix plays the same role as a translation operator. The eigenvalues of the transfer matrix construct a set of $e^{ika}~\left(k\in{\mathbb C}\right)$, which gives the generalized Brillouin zone. We note that, when there is a translation symmetry, eigenvalues of a transfer matrix belong to a set of $e^{ika}~\left(k\in{\mathbb R}\right)$~\cite{Dwivedi2016,Kunst2019}.

The explicit form of the transfer matrix in our setup can be obtained by
\begin{equation}
T={\cal P}\exp\left(i\frac{\omega}{c}\int_0^adzA\left(z\right)\right),
\label{eq5}
\end{equation}
where ${\cal P}$ is a path-ordered product. Importantly, we find that the generalized Brillouin zone is formed by the eigenvalues of the transfer matrix satisfying
\begin{equation}
\left|\rho_n\right|=\left|\rho_{n+1}\right|,
\label{eq6}
\end{equation}
where we number $2n$ eigenvalues so as to satisfy
\begin{equation}
\left|\rho_1\right|\leq\cdots\leq\left|\rho_{2n}\right|.
\label{eq7}
\end{equation}
The trajectories of $\beta=re^{i{\rm Re}\left(k\right)a}$ with $r\equiv\left|\rho_n\right|=\left|\rho_{n+1}\right|$ give the generalized Brillouin zone. Thereby, one can calculate the continuum bands by solving Eq.~(\ref{eq3}) with the complex-valued Bloch wavenumber. We note that Eq.~(\ref{eq6}) can be proved by using tight-binding models as shown in Appendix~\ref{secA}, and it has also been recently proved in continuous models~\cite{Hu2023}. In Sec.~\ref{sec3}, we shall demonstrate that the calculated continuum bands indeed reproduce the discrete eigenvalues of photonic crystals.

%
%

\section{\label{sec3}Photonic crystal}
Our main interest in this section lies in a chiral metamaterial which exhibits the magnetoelectric coupling via subwavelength structures. We then apply the non-Bloch band theory developed in Sec.~\ref{sec2} to photonic crystals composed of chiral metamaterials, which allows for investigating the non-Hermitian skin effect in the medium with linearly or circularly polarized eigenmodes.

%
%

\subsection{\label{sec3-1}Setup}
\begin{figure}[]
\includegraphics[width=7cm]{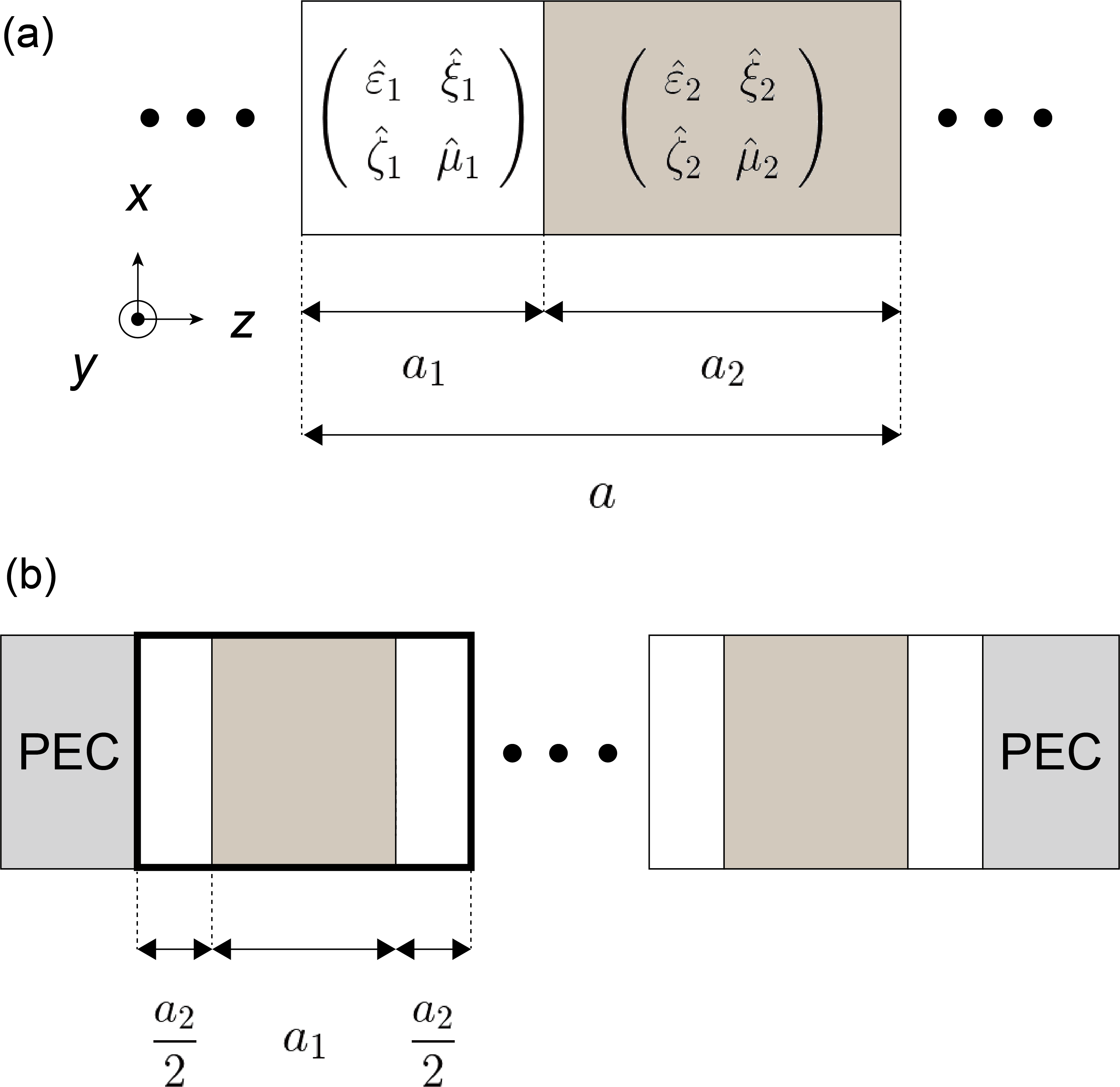}
\caption{\label{fig1}Schematic figures of the photonic crystal. (a) Unit cell to calculate the generalized Brillouin zone. The constitutive parameters and thickness of the layer $j~\left(j=1,2\right)$ are set to be $\left(\varepsilon_j,\xi_j,\zeta_j,\mu_j\right)$ and $a_j$, respectively. The lattice constant $a$ is given by $a_1+a_2$. (b) Finite-size photonic crystal. The black box represents the unit cell. We impose the perfect electric conductor (PEC) boundary condition on both ends of the system.}
\end{figure}
We consider the photonic crystal in which two media exhibiting the magnetoelectric coupling are alternately stacked in a periodic manner, where the layers 1 and 2 have the thicknesses $a_1$ and $a_2$, respectively [see Fig.~\ref{fig1}(a)]. We then investigate the eigenmodes of the photonic crystal surrounded by perfect electric conductors (PECs) without external excitation [see Fig.~\ref{fig1}(b)]. We note that the $x$ and $y$ components of the electric fields vanish at the ends of the system. Let us suppose that the electromagnetic waves form standing waves along the $z$ direction, and the polarization is perpendicular to the stacking direction. The electromagnetic waves in the photonic crystal can be described by the Maxwell's equations
\begin{eqnarray}
\left\{ \begin{array}{l}
{\bm\nabla}\times{\bm E}=i\omega{\bm B}, \vspace{5pt}\\
{\bm\nabla}\times{\bm H}=-i\omega{\bm D},
\end{array}\right.
\label{eq8}
\end{eqnarray}
associated with the constitutive relation at each layer
\begin{eqnarray}
\left( \begin{array}{c}
{\bm D} \vspace{5pt}\\
{\bm B}
\end{array}\right)=\left( \begin{array}{cc}
\varepsilon_0\hat{\varepsilon}_j & \hat{\xi}_j/c \vspace{5pt}\\
\hat{\zeta}_j/c                   & \mu_0\hat{\mu}_j
\end{array}\right)\left( \begin{array}{c}
{\bm E} \vspace{5pt}\\
{\bm H}
\end{array}\right),~\left(j=1,2\right),
\label{eq9}
\end{eqnarray}
where $\varepsilon_0$, $\mu_0$, and $c$ are the vacuum permittivity, the vacuum permeability, and the vacuum speed of light, respectively. Furthermore, $\hat{\varepsilon}_j$ and $\hat{\mu}_j$ are the relative permittivity and permeability tensors, respectively, and $\hat{\xi}_j$ and $\hat{\zeta}_j$ are the chirality tensors expressing the degrees of the magnetoelectric coupling. One can then rewrite the Maxwell's equations as follows:
\begin{eqnarray}
\frac{d}{dz}\left( \begin{array}{c}
\sqrt{\varepsilon_0}E_x\left(z\right) \vspace{5pt}\\
\sqrt{\varepsilon_0}E_y\left(z\right) \vspace{5pt}\\
\sqrt{\mu_0}H_x\left(z\right) \vspace{5pt}\\
\sqrt{\mu_0}H_y\left(z\right)
\end{array}\right)=i\frac{\omega}{c}A\left(z\right)\left( \begin{array}{c}
\sqrt{\varepsilon_0}E_x\left(z\right) \vspace{5pt}\\
\sqrt{\varepsilon_0}E_y\left(z\right) \vspace{5pt}\\
\sqrt{\mu_0}H_x\left(z\right) \vspace{5pt}\\
\sqrt{\mu_0}H_y\left(z\right)
\end{array}\right),
\label{eq10}
\end{eqnarray}
where $A\left(z\right)$ is represented by
\begin{eqnarray}
A\left(z\right)=i\left( \begin{array}{cc}
\sigma_y\hat{\zeta}_j        & \sigma_y\hat{\mu}_j  \vspace{5pt}\\
-\sigma_y\hat{\varepsilon}_j & -\sigma_y\hat{\xi}_j
\end{array}\right),
\label{eq11}
\end{eqnarray}
when $z$ is in the layer $j$. We note that $\sigma_0$ and $\left(\sigma_x,\sigma_y,\sigma_z\right)$ denote a $2\times2$ identity matrix and the Pauli matrices, respectively.

Throughout our paper, we study the case where the relative permittivity and permeability tensors satisfy $\hat{\varepsilon}_j=\varepsilon_j\sigma_0$ and $\hat{\mu}_j=\mu_j\sigma_0$, respectively, and $i\hat{\xi}_j\sigma_y$ and $-i\sigma_y\hat{\zeta}_j$ are simultaneously diagonalizable. Let $P_j$ denote the matrix diagonalizing $i\hat{\xi}_j\sigma_y$ and $-i\sigma_y\hat{\zeta}_j$. The linear transformations of the electric and magnetic fields at each layer,
\begin{eqnarray}
\begin{array}{l}
\left( \begin{array}{c}
\tilde{E}_+\left(z\right) \vspace{5pt}\\
\tilde{E}_-\left(z\right)
\end{array}\right)=P_j^{-1}\left( \begin{array}{c}
E_x\left(z\right) \vspace{5pt}\\
E_y\left(z\right)
\end{array}\right), \vspace{5pt}\\
\left( \begin{array}{c}
\tilde{H}_+\left(z\right) \vspace{5pt}\\
\tilde{H}_-\left(z\right)
\end{array}\right)=P_j^{-1}\sigma_y\left( \begin{array}{c}
H_x\left(z\right) \vspace{5pt}\\
H_y\left(z\right)
\end{array}\right),
\end{array}
\label{eq12}
\end{eqnarray}
then allow us to derive the block diagonalization form of Eq.~(\ref{eq10}) as follows:
\begin{widetext}
\begin{eqnarray}
\frac{d}{dz}\left( \begin{array}{c}
\sqrt{\varepsilon_0}\tilde{E}_+\left(z\right) \vspace{5pt}\\
\sqrt{\mu_0}\tilde{H}_+\left(z\right)         \vspace{5pt}\\
\sqrt{\varepsilon_0}\tilde{E}_-\left(z\right) \vspace{5pt}\\
\sqrt{\mu_0}\tilde{H}_-\left(z\right)
\end{array}\right)=i\frac{\omega}{c}\left( \begin{array}{cc}
\tilde{A}_+\left(z\right) & O                         \vspace{5pt}\\
O                         & \tilde{A}_-\left(z\right)
\end{array}\right)\left( \begin{array}{c}
\sqrt{\varepsilon_0}\tilde{E}_+\left(z\right) \vspace{5pt}\\
\sqrt{\mu_0}\tilde{H}_+\left(z\right)         \vspace{5pt}\\
\sqrt{\varepsilon_0}\tilde{E}_-\left(z\right) \vspace{5pt}\\
\sqrt{\mu_0}\tilde{H}_-\left(z\right)
\end{array}\right),
\label{eq13}
\end{eqnarray}
\end{widetext}
where $\tilde{A}_\pm\left(z\right)$, with $z$ being in the layer $j$, are represented by
\begin{eqnarray}
\tilde{A}_\pm\left(z\right)=\left( \begin{array}{cc}
-\zeta_{j,\pm}  & i\mu_j       \vspace{5pt}\\
-i\varepsilon_j & -\xi_{j,\pm}
\end{array}\right).
\label{eq14}
\end{eqnarray}
Here, $\xi_{j,\pm}$ and $\zeta_{j,\pm}$ are the eigenvalues of $i\hat{\xi}_j\sigma_y$ and $-i\sigma_y\hat{\zeta}_j$, respectively. It is worthwhile to mention that the eigenvectors of $i\hat{\xi}_j\sigma_y$ or, equivalently, $-i\sigma_y\hat{\zeta}_j$ determine polarizations of the eigenmodes. For the sake of later convenience, we denote the eigenvalues of Eq.~(\ref{eq14}) by $\lambda_{j,\pm},\bar{\lambda}_{j,\pm}$.

%
%

\subsection{\label{sec3-2}Transfer matrix and generalized Brillouin zone}
According to Eq.~(\ref{eq13}), the transfer matrix of the photonic crystal also has the block diagonalization form, which is given by
\begin{eqnarray}
T=\left( \begin{array}{cc}
T_+ & O   \vspace{5pt}\\
O   & T_-
\end{array}\right),
\label{eq15}
\end{eqnarray}
where
\begin{equation}
T_\pm={\cal P}\exp\left(i\frac{\omega}{c}\int_0^adz\tilde{A}_\pm\left(z\right)\right).
\label{eq16}
\end{equation}
To obtain the generalized Brillouin zone, we calculate the determinants of the block transfer matrices $T_\pm$. Let $U_{j,\pm}~\left(j=1,2\right)$ denote the matrices diagonalizing $T_\pm$. Equation~(\ref{eq16}) can then be written as
\begin{eqnarray}
&&T_\pm=U_{2,\pm}\left( \begin{array}{cc}
e^{i\omega\lambda_{2,\pm}a_2/c} & 0                                     \vspace{5pt}\\
0                               & e^{i\omega\bar{\lambda}_{2,\pm}a_2/c}
\end{array}\right)U_{2,\pm}^{-1} \nonumber\\
&&\times U_{1,\pm}\left( \begin{array}{cc}
e^{i\omega\lambda_{1,\pm}a_1/c} & 0                                     \vspace{5pt}\\
0                               & e^{i\omega\bar{\lambda}_{1,\pm}a_1/c}
\end{array}\right)U_{1,\pm}^{-1},
\label{eq17}
\end{eqnarray}
and one can obtain
\begin{equation}
\det T_\pm=\exp\left[i\frac{\omega}{c}\sum_{j=1}^2\left(\lambda_{j,\pm}+\bar{\lambda}_{j,\pm}\right)a_j\right].
\label{eq18}
\end{equation}

We note that, in this case, the following facts enable us to straightforwardly calculate the generalized Brillouin zone. First, each of the eigenvalues $\rho_{1,\pm},\rho_{2,\pm}$ of the block transfer matrices $T_\pm$ induces the condition for the generalized Brillouin zone, which means that
\begin{equation}
\left|\rho_{1,\pm}\right|=\left|\rho_{2,\pm}\right|.
\label{eq19}
\end{equation}
Second, the product of the eigenvalues is equal to the determinant of the block transfer matrix. Therefore, one can get the generalized Brillouin zones as closed loops formed by $r_\pm e^{i{\rm Re}\left(k\right)a}$, where $r_\pm$ are given by
\begin{equation}
r_\pm=\sqrt{\left|\det T_\pm\right|}.
\label{eq20}
\end{equation}

%
%

\subsection{\label{sec3-3}Linearly polarized eigenmodes}
We first investigate the case where the eigenmodes of the photonic crystal are linearly polarized. The chirality tensors are then given by
\begin{eqnarray}
\hat{\xi}_j=\left( \begin{array}{cc}
0          & \xi_{xy,j} \vspace{5pt}\\
\xi_{yx,j} & 0
\end{array}\right),~\hat{\zeta}_j=\left( \begin{array}{cc}
0            & \zeta_{xy,j} \vspace{5pt}\\
\zeta_{yx,j} & 0
\end{array}\right)
\label{eq21}
\end{eqnarray}
with $j=1,2$. One can easily check that the eigenvectors of $i\hat{\xi}_j\sigma_y$ and $-i\sigma_y\hat{\zeta}_j$ are indeed linearly polarized, i.e., they are proportional to $\left(1,0\right)^{\rm T}$ and $\left(0,1\right)^{\rm T}$. The corresponding eigenvalues are given by $\left(\xi_{j,+},\xi_{j,-}\right)=\left(-\xi_{xy,j},\xi_{yx,j}\right)$ and $\left(\zeta_{j,+},\zeta_{j,-}\right)=\left(-\zeta_{yx,j},\zeta_{xy,j}\right)$. We focus on the governing equation obtained from one block of Eq.~(\ref{eq13}), which can be written as
\begin{eqnarray}
\frac{d}{dz}\left( \begin{array}{c}
\sqrt{\varepsilon_0}E_x\left(z\right) \vspace{5pt}\\
\sqrt{\mu_0}H_y\left(z\right)
\end{array}\right)=i\frac{\omega}{c}A\left(z\right)\left( \begin{array}{c}
\sqrt{\varepsilon_0}E_x\left(z\right) \vspace{5pt}\\
\sqrt{\mu_0}H_y\left(z\right)
\end{array}\right),
\label{eq22}
\end{eqnarray}
where $A\left(z\right)$ is represented by
\begin{eqnarray}
A\left(z\right)=\left( \begin{array}{cc}
\zeta_{yx,j}  & \mu_j      \vspace{5pt}\\
\varepsilon_j & \xi_{xy,j}
\end{array}\right),
\label{eq23}
\end{eqnarray}
when $z$ is in the layer $j$. We note that Eq.~(\ref{eq22}) and the other block of Eq.~(\ref{eq13}) describe the linearly polarized eigenmodes whose polarization vectors lie in the $x$ and $y$ axis, respectively. Clearly, these eigenmodes are independent of each other.

As shown in Sec.~\ref{sec3-2}, the generalized Brillouin zone can be obtained by $re^{i{\rm Re}\left(k\right)a}$, where $r$ is given by
\begin{equation}
r=\exp\left[-\frac{1}{2c}{\rm Im}\left(\omega\sum_{j=1}^2\left(\xi_{xy,j}+\zeta_{yx,j}\right)a_j\right)\right].
\label{eq24}
\end{equation}
We note that Eq.~(\ref{eq24}) reproduces the result obtained in Ref.~\cite{Yan2021} when $\xi_{xy,j}=\zeta_{yx,j}$. One can infer from Eq.~(\ref{eq24}) that not only the chirality parameters but also the eigenfrequencies contribute to the localization of the non-Hermitian skin effect. Meanwhile, the non-Hermitian skin effect disappears when the system becomes reciprocal, i.e., $\hat{\xi}=-\hat{\zeta}^{\rm T}$.

\begin{figure}[]
\includegraphics[width=8.5cm]{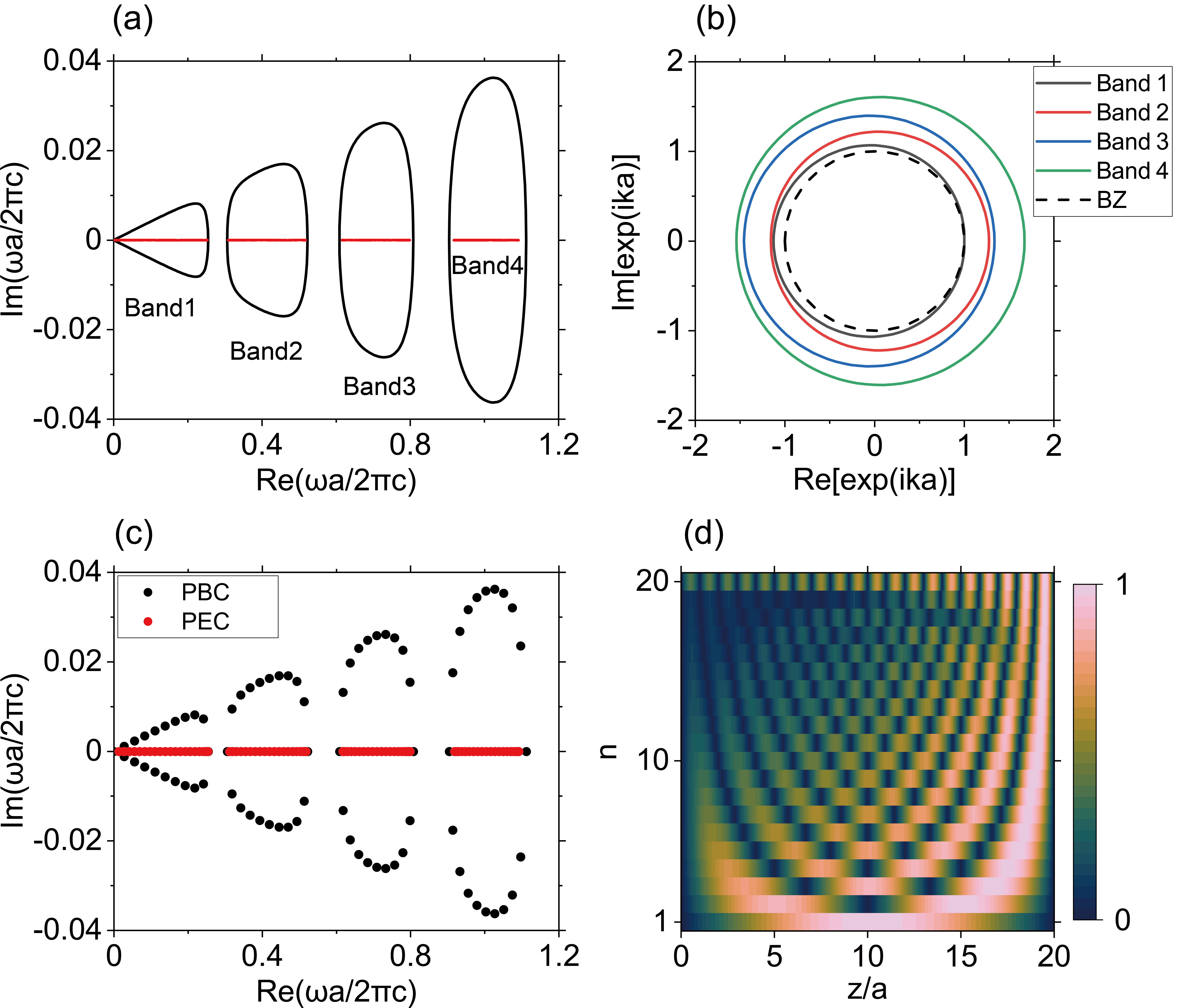}
\caption{\label{fig2}Non-Hermitian skin effect of the linearly polarized eigenmodes. (a) Continuum bands calculated from the complex-valued Bloch wavenumber (red) and real-valued Bloch wavenumber (black). (b) Generalized Brillouin zones (color loops) and conventional Brillouin zone (black dashed circle). (c) Eigenvalues of a finite-size system under the perfect electric conductor (PEC) boundary condition (red) and a periodic boundary condition (PBC) (black). The system size is set to be $20a$. (d) Spatial profiles of all eigenmodes included in band 1. We set the parameters as $\varepsilon_1=\mu_1=1,\xi_{xy,1}=\zeta_{yz,1}=0,\varepsilon_2=4,\mu_2=1,\xi_{xy,2}=-0.3i,\zeta_{yx,2}=0.1$, and $a_1=0.75a$.}
\end{figure}

Equation~(\ref{eq24}) allows us to determine the generalized Brillouin zones [see Fig.~\ref{fig2}(b)], which deviate from the conventional Brillouin zone. We then calculate the continuum bands from the generalized Brillouin zones [red lines in Fig.~\ref{fig2}(a)] and confirm that the continuum bands reproduce the eigenvalues of a finite-size system [red dots in Fig.~\ref{fig2}(c)]. We note that the red continuum bands are distinct from the ones obtained from the real-valued Bloch wave number [black curves in Fig.~\ref{fig2}(a)]. Finally, Fig.~\ref{fig2}(d) shows that the eigenmodes are localized due to the non-Hermitian skin effect, and this localization occurs at only one side of the system.

%
%

\subsection{\label{sec3-4}Circularly polarized eigenmodes}
We next investigate the case where the eigenmodes of the photonic crystal are circularly polarized. The chirality tensors are given by
\begin{equation}
\hat{\xi}_j=\xi_j\sigma_0,~\hat{\zeta}_j=\zeta_j\sigma_0,~\left(j=1,2\right),
\label{eq25}
\end{equation}
for which the eigenvectors of $i\hat{\xi}_j\sigma_y$ and $-i\sigma_y\hat{\zeta}_j$ are proportional to $\left(1,\pm i\right)^{\rm T}$. The corresponding eigenvalues are given by $\left(\xi_{j,+},\xi_{j,-}\right)=\left(-i\xi_j,i\xi_j\right)$ and $\left(\zeta_{j,+},\zeta_{j,-}\right)=\left(i\zeta_j,-i\zeta_j\right)$. The governing equations then read
\begin{eqnarray}
\frac{d}{dz}\left( \begin{array}{c}
\sqrt{\varepsilon_0}E_\pm\left(z\right) \vspace{5pt}\\
\sqrt{\mu_0}H_\pm\left(z\right)
\end{array}\right)=i\frac{\omega}{c}A_\pm\left(z\right)\left( \begin{array}{c}
\sqrt{\varepsilon_0}E_\pm\left(z\right) \vspace{5pt}\\
\sqrt{\mu_0}H_\pm\left(z\right)
\end{array}\right),
\label{eq26}
\end{eqnarray}
where $A_\pm\left(z\right)$ is represented by
\begin{eqnarray}
A_\pm\left(z\right)=\left( \begin{array}{cc}
\mp i\zeta_j       & i\mu_j \vspace{5pt}\\
-i\varepsilon_j    & \pm i\xi_j
\end{array}\right)
\label{eq27}
\end{eqnarray}
when $z$ is in the layer $j$. We note that $\left(E_+\left(z\right),H_+\left(z\right)\right)$ and $\left(E_-\left(z\right),H_-\left(z\right)\right)$ correspond to the left- and right-circularly polarized eigenmodes, respectively. In contrast to the linearly polarized case discussed in Sec.~\ref{sec3-3}, these circularly polarized eigenmodes are not independent of each other, which means that the corresponding eigenvalues are degenerate. Equation~(\ref{eq20}) allows us to calculate the generalized Brillouin zones of the left- $\left(+\right)$ and right- $\left(-\right)$ circularly polarized eigenmodes, where $r_\pm$ reads
\begin{equation}
r_+=\exp\left[-\frac{1}{2c}{\rm Re}\left(\omega\sum_{j=1}^2\left(\xi_j-\zeta_j\right)a_j\right)\right],~r_-=\frac{1}{r_+}.
\label{eq28}
\end{equation}

\begin{figure}[]
\includegraphics[width=8.5cm]{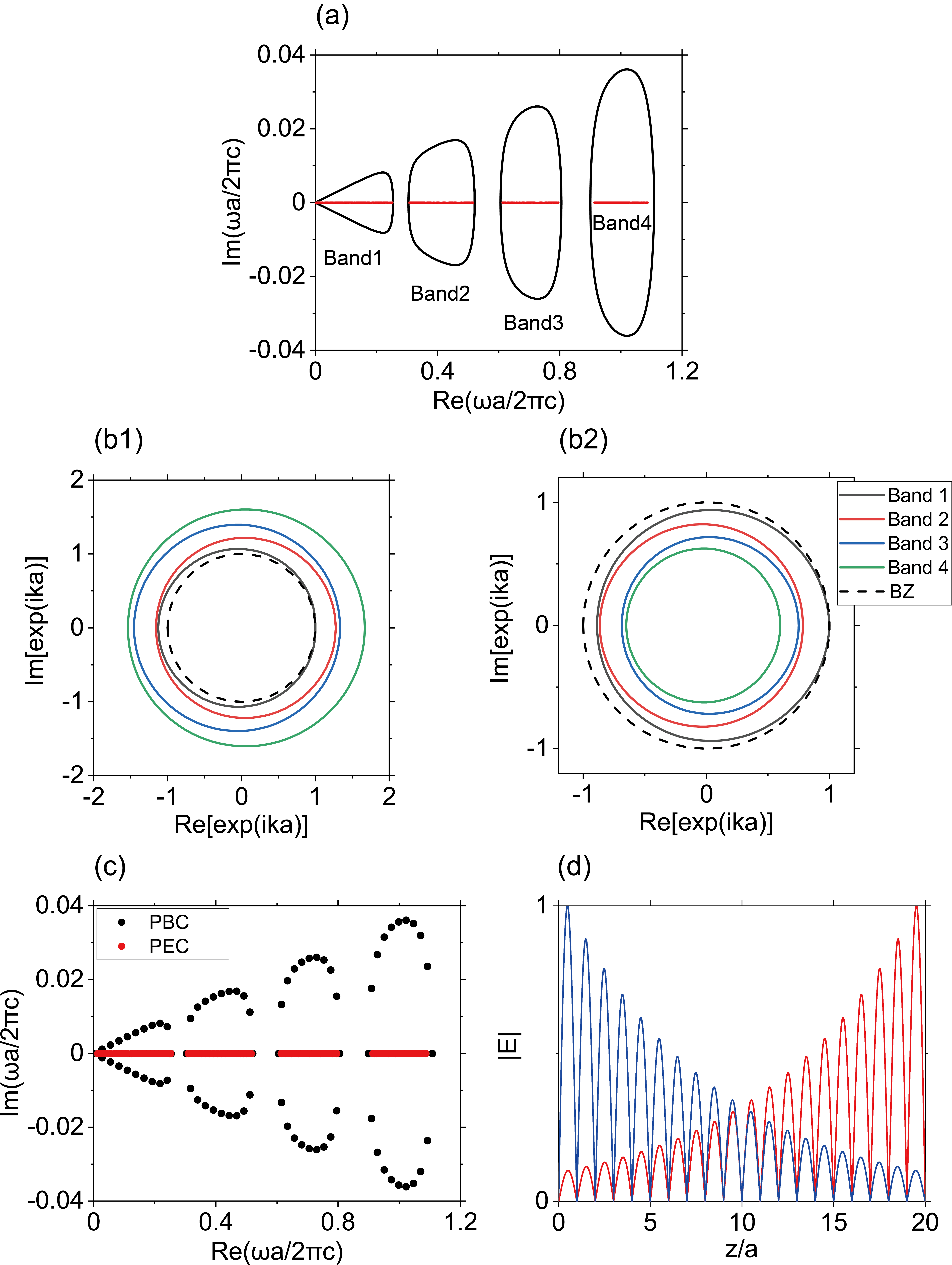}
\caption{\label{fig3}Non-Hermitian skin effect of the circularly polarized eigenmodes. (a) Continuum bands calculated from the complex-valued Bloch wave number (red) and real-valued Bloch wave number (black). (b) Generalized Brillouin zones (color loops) and conventional Brillouin zone (black dashed loop). The generalized Brillouin zones of the left- and right-circularly polarized eigenmodes are shown in (b1) and (b2), respectively. (c) Eigenvalues of a finite-size system under the perfect electric conductor (PEC) boundary condition (red) and a periodic boundary condition (PBC) (black). The system size is set to be $20a$. (d) Spatial profiles of the left- (red) and right- (blue) circularly polarized eigenmodes with $\omega a/2\pi c=0.2523$. We set the parameters as $\varepsilon_1=\mu_1=1,\xi_{xy,1}=\zeta_{yz,1}=0,\varepsilon_2=4,\mu_2=1,\xi_2=0.1,\zeta_2=-0.1$, and $a_1=0.75a$.}
\end{figure}

We show the generalized Brillouin zones of the left- and right-circularly polarized eigenmodes in Figs.~\ref{fig3}(b1) and (b2), respectively, and the corresponding continuum bands are shown in Fig.~\ref{fig3}(a). One can confirm that the eigenvalues of the two circularly polarized eigenmodes are degenerate, while the generalized Brillouin zones are distinct from each other. Figures.~\ref{fig3}(a) and \ref{fig3}(c) indicate that the continuum bands indeed reproduce the eigenvalues of a finite-size system. Furthermore, as shown in Fig.~\ref{fig3}(d), the left-circularly polarized eigenmode (red) is localized at the right boundary, while the right-circularly polarized eigenmode (blue) is localized at the opposite boundary. These localization behaviors are in a stark contrast to the case of the medium with the linearly polarized eigenmodes; in the latter, all the eigenmodes localize to the edge of the same side. We emphasize that this qualitative difference originates from the reciprocity operation defined in Ref.~\cite{Yan2021}. Indeed, the medium considered here is invariant under the reciprocity operation, which converts the right- (left-) circularly polarized eigenmodes to the left (right) ones. Thereby, the non-Hermitian skin effect persists even if the system is reciprocal, i.e., $\hat{\xi}=-\hat{\zeta}^{\rm T}$.

%
%

\section{\label{sec4}Summary and Discussions}
In this paper, we develop the non-Bloch band theory of non-Hermitian systems described by the generalized eigenvalue equation. One can calculate the generalized Brillouin zone from the eigenvalues of the transfer matrix. The theory allows us to investigate non-Hermitian physics of multicomponent wave functions exhibiting the non-Hermitian skin effect, in contrast to Refs.~\cite{Yokomizo2022,Hu2023} which have studied the non-Hermitian skin effect of single-component wave functions. Indeed, we demonstrate that the theory is applicable to electromagnetic waves in the photonic crystals composed of the chiral metamaterials. As a result, we find that the localization properties of the skin modes can qualitatively change depending on the polarization of the eigenmodes. Notably, when the medium has circularly polarized eigenmodes, the skin modes can localize at either of edges depending on whether it is left- or right-circularly polarized.

We remark that there is plenty of room to extend the theoretical analysis performed in the present paper. While we have focused on the case where the medium has linearly or circularly polarized eigenmodes, one can study the non-Hermitian skin effect of elliptically polarized eigenmodes by setting the chirality tensors to be
\begin{eqnarray}
\hat{\xi}_j=\left( \begin{array}{cc}
\xi_{xx,j} & 0          \vspace{5pt}\\
0          & \xi_{yy,j}
\end{array}\right),~\hat{\zeta}_j=\left( \begin{array}{cc}
\zeta_{xx,j} & 0            \vspace{5pt}\\
0            & \zeta_{yy,j}
\end{array}\right),
\label{eq29}
\end{eqnarray}
satisfying $\xi_{xx,j}\zeta_{xx,j}=\xi_{yy,j}\zeta_{yy,j}$ with $j=1,2$. Additionally, the non-Hermitian skin effect of various polarized eigenmodes can be explored by changing the forms of $\hat{\varepsilon},\hat{\xi},\hat{\zeta}$, and $\hat{\mu}$, instead of the present setup. We expect that one can study topological edge states of the photonic crystal by defining topological invariants from the generalized Brillouin zone, similar to the way discussed in Refs.~\cite{Yokomizo2022,Yan2021}.

It merits further study to analyze how the eigenmodes of the photonic crystal with open boundary conditions can actually be excited by incident light. We point out that the results revealed in this paper suggest the possibility of controlling the localization properties of excitation modes simply by changing a polarization of incident light while keeping the medium unchanged. Indeed, there has been an experimental proposal to modulate excitation modes by circularly polarized incident light~\cite{Wang2019}. For example, in the photonic crystal investigated in Sec.~\ref{sec3-4}, we expect that a modulation of incident light from a linear polarization to a circular polarization would change the localization positions of the excitation modes from both ends to one end. 

In practice, photonic crystals have complex-valued eigenfrequencies because optical gain and loss are unbalanced, unlike the photonic crystal considered in this paper. Our previous work has investigated the excitation of the corresponding eigenmodes by incident light with real-valued frequencies~\cite{Yoda2023}. Meanwhile, it is intriguing to study the localization behavior of those excited modes. Indeed, the non-Hermitian skin effect in the transient regime has been experimentally observed in acoustics~\cite{Gu2022}. The investigation of the transient non-Hermitian skin effect in photonics is left as a future work.

Finally, we comment on the feasibility of the photonic crystals and the measurement possibility of the skin modes. We remark that Refs.~\cite{Zhao2010,Droulias2019} proposed a chiral metamaterial slab which has circularly polarized eigenmodes. Hence, the periodic arrangement of the chiral metamaterial would realize the photonic crystal considered in Sec.~\ref{sec3-4}. Meanwhile, to our knowledge, there have been no proposals for chiral metamaterials which have linearly polarized eigenmodes. Nevertheless, multiferroic materials would allow us to realize the photonic crystal considered in Sec.~\ref{sec3-3}. Furthermore, it would be possible to observe the skin modes emerging in the photonic crystals in the same way as the measurements of topological edge states, directly \cite{Parappurath2020} or indirectly \cite{Huang2019}.

%
%

\begin{acknowledgements}
We are grateful to S. Murakami, R. Okugawa, R. Takahashi, and T. Yoshida for variable discussion. K. Y. was supported by JSPS KAKENHI (Grants No. JP21J01409 and No. JP23K13027). Y. A. acknowledges support from the Japan Society for the Promotion of Science (JSPS) through Grant No. JP19K23424 and from JST FOREST Program (Grant No. JPMJFR222U, Japan) and JST CREST (Grant No. JPMJCR2312, Japan).
\end{acknowledgements}

%
%

\appendix

%
%

\section{\label{secA}Transfer matrix in tight-binding models}
In this appendix, we discuss a method to construct a transfer matrix in 1D non-Hermitian tight-binding systems along the lines of Ref.~\cite{Kunst2019}. We derive the condition for the generalized Brillouin zone in terms of the eigenvalues of the transfer matrix. One can indeed see that the band structure calculated from the generalized Brillouin zone reproduces the energy eigenvalues under open boundary conditions.

%
%

\subsection{\label{secA-1}Setup}
We start from 1D non-Hermitian tight-binding systems which are described by
\begin{equation}
H=\sum_i\sum_{j=-N}^N\sum_{\mu,\nu=1}^qt_{j,\mu\nu}c_{i+j,\mu}^\dag c_{i,\nu},
\label{eqappa1}
\end{equation}
where $c_{i,\nu}$ is an annihilation operator of a particle at sublattice $\nu$ in the $i$th unit cell, and $t_{j,\mu\nu}$ is not equal to $t_{-j,\nu\mu}^\ast$. We assume that the particles hop to the $N$th-nearest unit cell. In Eq.~(\ref{eqappa1}), sublattice sites in a unit cell are connected to up to the $2N$th-nearest-neighbor unit cells via the hopping amplitudes. Meanwhile, we can enlarge a unit cell so sublattice sites in a given cell are connected to only the nearest-neighbor cells~\cite{Dwivedi2016,Kunst2019}. The enlarged unit cell is called a supercell, including $s\geq qN$ degrees of freedom, the definition of which is not unique. Let ${\bm c}_n$ denote a vector of the annihilation operators of the particles in the $n$th supercell. Equation~(\ref{eqappa1}) can then be rewritten as the reduced form including only the nearest-neighbor hopping amplitudes,
\begin{equation}
H=\sum_n\left({\bm c}^\dag_nJ_L{\bm c}_{n+1}+{\bm c}_n^\dag M{\bm c}_n+{\bm c}_{n+1}^\dag J_R^\dag{\bm c}_n\right),
\label{eqappa2}
\end{equation}
where $J_L$ and $J_R$ are hopping matrices, and $M$ is an on-site matrix. The reduced form of the real-space eigenequation reads
\begin{equation}
J_L{\bm \Psi}_{n+1}+M{\bm \Psi}_n+J_R^\dag{\bm \Psi}_{n-1}=E{\bm \Psi}_n,
\label{eqappa3}
\end{equation}
where ${\bm \Psi}_n$ is a wave function for the $n$th supercell. We note that the systems become Hermitian when $J_L=J_R$ and $M$ is a Hermitian matrix.

In the following, we focus on the systems with bidirectional hopping amplitudes, e.g., the non-Hermitian Su-Schrieffer-Heeger (SSH) model [see Fig.~\ref{figappa1}(a)]. The hopping matrices $J_L$ and $J_R$ are then written in the form of a lower or upper triangular matrix with zero diagonal elements, satisfying
\begin{equation}
J_L^2=J_R^2=O.
\label{eqappa4}
\end{equation}
We note that taking a sufficiently large supercell ensures Eq.~(\ref{eqappa4}). Indeed, the physical interpretation of Eq.~(\ref{eqappa4}) is that, in a given supercell, no sublattice sites are connected to both the left and right adjacent supercells. For the sake of simplicity, we further assume that
\begin{equation}
{\rm rank}\left(J_L\right)={\rm rank}\left(J_R\right)=r,
\label{eqappa5}
\end{equation}
which means that, in a given supercell, $r$ sublattice sites are connected to the adjacent supercell. In that situation, we derive the transfer matrix of the systems from Eq.~(\ref{eqappa3}). To this end, we utilize the singular value decompositions of the hopping matrices given by
\begin{eqnarray}
\left\{ \begin{array}{l}
J_L=X\Xi_LY^\dag, \vspace{5pt}\\
J_R=V\Xi_RW^\dag,
\end{array}\right.
\label{eqappa6}
\end{eqnarray}
where $\Xi_L$ and $\Xi_R$ are $r\times r$ diagonal matrices including the singular values of $J_L$ and $J_R$, respectively, and $X,Y,V$, and $W$ are $s\times r$ matrices satisfying
\begin{equation}
X^\dag X=Y^\dag Y=V^\dag V=W^\dag W=\1,
\label{eqappa7}
\end{equation}
and
\begin{equation}
X^\dag Y=V^\dag W=O.
\label{eqappa8}
\end{equation}
It is worth noting that Eq.~(\ref{eqappa8}) is always ensured because of Eq.~(\ref{eqappa4}). We can then rewrite Eq.~(\ref{eqappa3}) as
\begin{equation}
{\bm\Psi}_n=GX\Xi_L{\bm\beta}_{n+1}+GW\Xi_R{\bm\alpha}_{n-1},
\label{eqappa9}
\end{equation}
where ${\bm\alpha}_n=V^\dag{\bm\Psi}_n$, ${\bm\beta}_n=Y^\dag{\bm\Psi}_n$, and $G=\left(E-M\right)^{-1}$. For the sake of convenience, let $G_{ab}$ denote $B^\dag GA$ for $s\times r$ matrices $A,B$. Equation.~(\ref{eqappa9}) leads to the recursion equation for ${\bm\alpha}_n$ and ${\bm\beta}_n$ given by
\begin{eqnarray}
\left( \begin{array}{c}
{\bm\beta}_{n+1} \vspace{5pt}\\
{\bm\alpha}_n
\end{array}\right)=T\left( \begin{array}{c}
{\bm\beta}_n \vspace{5pt}\\
{\bm\alpha}_{n-1}
\end{array}\right),
\label{eqappa10}
\end{eqnarray}
where $T$ is the transfer matrix explicitly written as
\begin{eqnarray}
T=\left( \begin{array}{cc}
\Xi_L^{-1}G_{xy}^{-1} & -\Xi_LG_{xy}^{-1}G_{wy}\Xi_R                     \vspace{5pt}\\
G_{xv}G_{xy}^{-1}     & \left(G_{wv}-G_{xv}G_{xy}^{-1}G_{wy}\right)\Xi_R
\end{array}\right).
\label{eqappa11}
\end{eqnarray}
We note that the size of the transfer matrix is independent of the choice of a supercell.

%
%

\subsection{\label{secA-2}Condition for the generalized Brillouin zone}
We show that the condition for the generalized Brillouin zone can be obtained from the eigenvalues of the transfer matrix. For the sake of simplicity, we suppose that the transfer matrix is a diagonalizable matrix, for which the eigenvalue problem is written as
\begin{equation}
T{\bm\varphi}^{\left(l\right)}=\rho_l{\bm\varphi}^{\left(l\right)}
\label{eqapp12}
\end{equation}
with $l=1,\cdots,2r$, where we number the eigenvalues so as to satisfy $\left|\rho_1\right|\leq\cdots\leq\left|\rho_{2r}\right|$. In the following, we focus on a finite-size system with open boundary conditions expressed by
\begin{equation}
{\bm\Psi}_0={\bm\Psi}_{L+1}=0,
\label{eqappa13}
\end{equation}
where $L$ is a system size. One can immediately obtain the boundary equation from Eq.~(\ref{eqappa10}) as follows:
\begin{eqnarray}
\left( \begin{array}{c}
{\bm0}        \vspace{5pt}\\
{\bm\alpha}_L
\end{array}\right)=T^L\left( \begin{array}{c}
{\bm\beta}_1 \vspace{5pt}\\
{\bm0}
\end{array}\right).
\label{eqappa14}
\end{eqnarray}
We note that one can take ${\bm\alpha}_L$ and ${\bm\beta}_1$ to be arbitrary vectors. To solve Eq.~(\ref{eqappa14}), we expand the vectors included in this equation in terms of the eigenvectors of the transfer matrix, which leads to
\begin{eqnarray}
\left( \begin{array}{c}
{\bm\beta}_1 \vspace{5pt}\\
{\bm0}
\end{array}\right)&=&\sum_{l=1}^{2r}a_l{\bm\varphi}^{\left(l\right)}, \label{eqappa15}\\
\left( \begin{array}{c}
{\bm0}        \vspace{5pt}\\
{\bm\alpha}_L
\end{array}\right)&=&\sum_{l=1}^{2r}a_l\left(\rho_l\right)^L{\bm\varphi}^{\left(l\right)}.
\label{eqappa16}
\end{eqnarray}
Let ${\cal P}_\alpha=\left(O,\1\right)$ and ${\cal P}_\beta=\left(\1,O\right)$ denote $r\times2r$ matrices. Acting ${\cal P}_\alpha$ and ${\cal P}_\beta$ on Eqs.~(\ref{eqappa15}) and (\ref{eqappa16}), respectively, we get
\begin{eqnarray}
\begin{array}{l}
\displaystyle\sum_{l=1}^{2r}a_l{\cal P}_\alpha{\bm\varphi}^{\left(l\right)}={\bm0},                     \vspace{5pt}\\
\displaystyle\sum_{l=1}^{2r}a_l\left(\rho_l\right)^L{\cal P}_\beta{\bm\varphi}^{\left(l\right)}={\bm0}.
\end{array}
\label{eqappa17}
\end{eqnarray}
Equation~(\ref{eqappa17}) is a set of algebraic equations for $a_1,\cdots,a_{2r}$, and we can recast it to the form of a matrix equation as follows:
\begin{eqnarray}
\left(R_1{\bm\varphi}^{\left(1\right)}~\cdots~R_{2r}{\bm\varphi}^{\left(2r\right)}\right)\left( \begin{array}{c}
a_1    \vspace{5pt}\\
\vdots \vspace{5pt}\\
a_{2r}
\end{array}\right)={\bm0},
\label{eqappa18}
\end{eqnarray}
where $R_l$ are $2r\times2r$ matrices defined by
\begin{eqnarray}
R_l=\left( \begin{array}{cc}
\left(\rho_l\right)^L\1 & O  \vspace{5pt}\\
O                       & \1
\end{array}\right).
\label{eqappa19}
\end{eqnarray}
The condition that Eq.~(\ref{eqappa18}) has a nontrivial solution finally reads
\begin{eqnarray}
\left| \begin{array}{ccc}
\varphi_1^{\left(1\right)}\left(\rho_1\right)^L & \cdots & \varphi_1^{\left(2r\right)}\left(\rho_{2r}\right)^L \vspace{5pt}\\
\vdots                                          & \vdots & \vdots                                              \vspace{5pt}\\
\varphi_r^{\left(1\right)}\left(\rho_1\right)^L & \cdots & \varphi_r^{\left(2r\right)}\left(\rho_{2r}\right)^L \vspace{5pt}\\
\varphi_{r+1}^{\left(1\right)}                  & \cdots & \varphi_{r+1}^{\left(2r\right)}                     \vspace{5pt}\\
\vdots                                          & \vdots & \vdots                                              \vspace{5pt}\\
\varphi_{2r}^{\left(1\right)}                   & \cdots & \varphi_{2r}^{\left(2r\right)}
\end{array}\right|=0.
\label{eqappa20}
\end{eqnarray}
We remark that one can get the eigenenergies of a finite-size system with edges by solving Eq.~(\ref{eqappa20}), since this equation is defined only by the eigenvalues and eigenvectors of the transfer matrix.

To study the condition for the generalized Brillouin zone, we expand Eq.~(\ref{eqappa20}) as follows:
\begin{equation}
\sum_{P,P^\prime}\sum_{Q,Q^\prime}F\left(\varphi_{i\in Q}^{\left(j\in P\right)},\varphi_{i^\prime\in Q^\prime}^{\left(j^\prime\in P^\prime\right)}\right)\prod_{i\in P}\left(\rho_i\right)^L=0,
\label{eqappa21}
\end{equation}
where $P$ and $P^\prime$ denote two disjoint subsets of the set $\left\{1,\cdots,2r\right\}$ with $r$ elements, $Q=\left\{1,\cdots,r\right\}$, and $Q^\prime=\left\{r+1,\cdots,2r\right\}$. While it is difficult to get the exact solutions of Eq.~(\ref{eqappa21}), in general, one can investigate the asymptotic behavior of the eigenenergies in $L\rightarrow\infty$. The key observation is that, when
\begin{equation}
\left|\rho_r\right|=\left|\rho_{r+1}\right|
\label{eqappa22}
\end{equation}
is satisfied, the dominant contributions to the left-hand side of Eq.~(\ref{eqappa21}) are the term including $\left(\rho_r\rho_{r+2}\cdots\rho_{2r}\right)^L$ and the one including $\left(\rho_{r+1}\rho_{r+2}\cdots\rho_{2r}\right)^L$ in the limit of a large system size. As a result, the asymptotic form of Eq.~(\ref{eqappa21}) can be obtained by
\begin{equation}
\left(\frac{\rho_r}{\rho_{r+1}}\right)^L=-\frac{F\left(\varphi_{i\in Q}^{\left(j\in P_0\right)},\varphi_{i^\prime\in Q^\prime}^{\left(j^\prime\in P_0^\prime\right)}\right)}{F\left(\varphi_{i\in Q}^{\left(j\in P_1\right)},\varphi_{i^\prime\in Q^\prime}^{\left(j^\prime\in P_1^\prime\right)}\right)},
\label{eqappa23}
\end{equation}
where $P_0=\left\{r+1,\cdots,2r\right\},P_0^\prime=\left\{1,\cdots,r\right\},P_1=\left\{r,r+2,\cdots,2r\right\}$, and $P_1^\prime=\left\{1,\cdots,r-1,r\right\}$. Thus, one can get the continuum energy bands, since the change of the relative phase between $\rho_r$ and $\rho_{r+1}$ produces dense sets of the eigenenergies. Accordingly, the sets of the eigenvalues of the transfer matrix satisfying Eq.~(\ref{eqappa22}) form the generalized Brillouin zone.

%
%

\subsection{\label{secA-3}Non-Hermitian Su-Schrieffer-Heeger model}
\begin{figure}[]
\includegraphics[width=8.5cm]{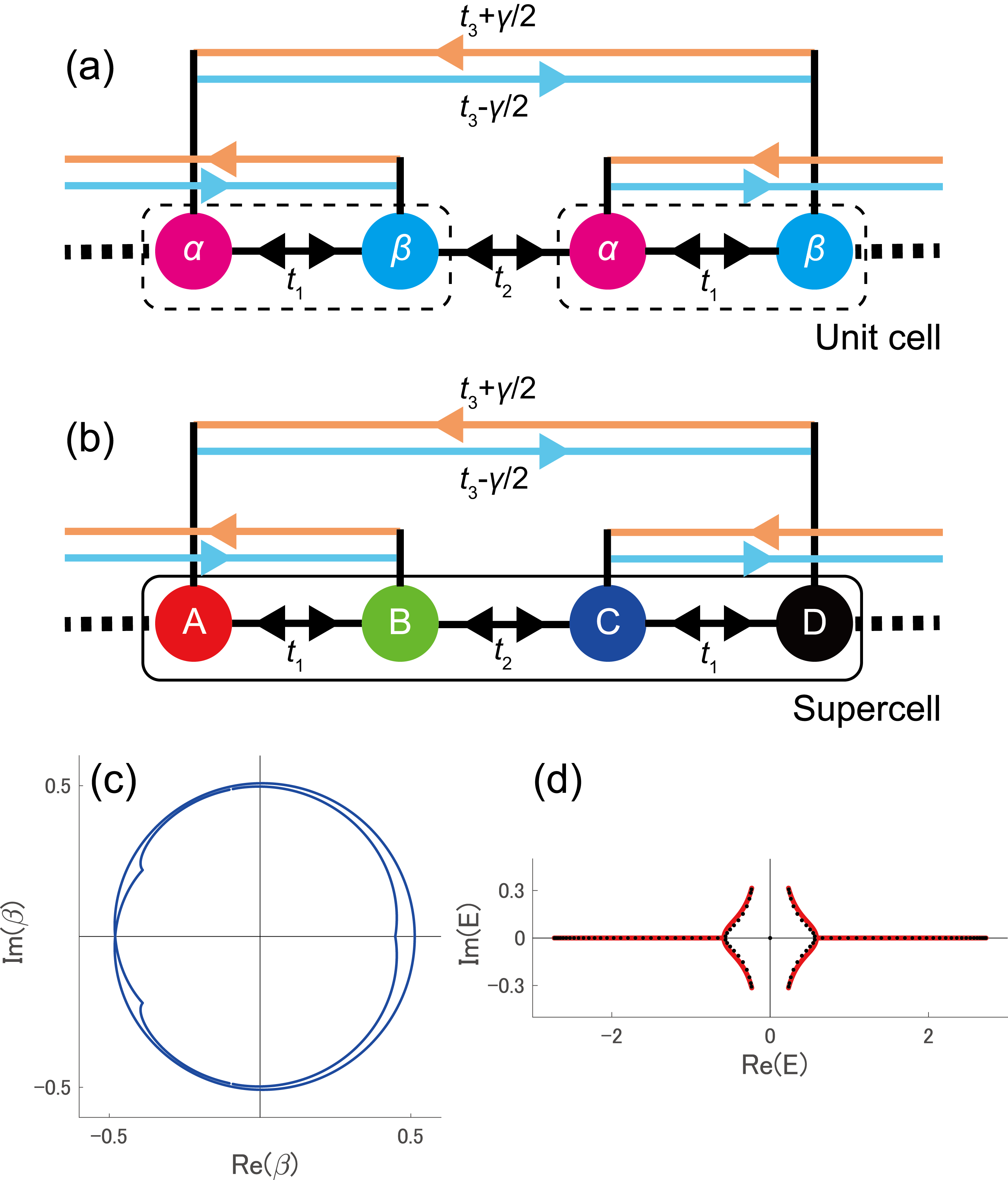}
\caption{\label{figappa1}Schematic figures, generalized Brillouin zones, and eigenenergies of the non-Hermitian Su-Schrieffer-Heeger model. (a) Unit cells represented by the dashed boxes. (b) Supercell of our choice represented by the box. (c) Generalized Brillouin zones. (d) Continuum energy bands (red) and eigenenergies of a finite-size system with open boundary conditions (black). The system size is set to be $L=50$. We set the system parameters as $t_1=1,t_2=0.7,t_3=1.2$, and $\gamma=4/3$.}
\end{figure}

We show that Eq.~(\ref{eqappa22}) allows us to calculate the generalized Brillouin zone and the continuum energy bands by using the non-Hermitian SSH model with asymmetric next-nearest-neighbor hopping amplitudes [Fig.~\ref{figappa1}(a)]. The real-space Hamiltonian of this system reads
\begin{eqnarray}
H&=&\sum_n\left[t_1\left(c_{n,\alpha}^\dag c_{n,\beta}+c_{n,\beta}^\dag c_{n,\alpha}\right)\right. \nonumber\\
&&+t_2\left(c_{n,\beta}^\dag c_{n+1,\alpha}+c_{n+1,\alpha}^\dag c_{n,\beta}\right) \nonumber\\
&&\left. +\left(t_3+\frac{\gamma}{2}\right)c_{n,\alpha}^\dag c_{n+1,\beta}+\left(t_3-\frac{\gamma}{2}\right)c_{n+1,\beta}^\dag c_{n,\alpha}\right], \nonumber\\
\label{eqappa24}
\end{eqnarray}
where all the parameters take positive values, and we assume $t_3>\gamma/2$ for simplicity. We note that Eq.~(\ref{eqappa24}) has the form of Eq.~(\ref{eqappa1}) with $N=1$ and $q=2$.

To calculate the transfer matrix of the system in the manner explained in Sec.~\ref{secA-1}, we take a supercell including four sublattices [Fig.~\ref{figappa1}(b)]. One can then obtain the reduced Hamiltonian in the form of Eq.~(\ref{eqappa2}) and also the reduced real-space eigenequation for the wavefunction ${\bm\Psi}_n=\left(\Psi_{n,{\rm A}},\Psi_{n,{\rm B}},\Psi_{n,{\rm C}},\Psi_{n,{\rm D}}\right)$ in the form of Eq.~(\ref{eqappa3}). Here, the hopping matrices and onsite matrix are given by
\begin{eqnarray}
J_R^\dag&=&\left( \begin{array}{cccc}
0 & 0 & 0            & t_2 \vspace{5pt}\\
0 & 0 & t_3-\gamma/2 & 0   \vspace{5pt}\\
0 & 0 & 0            & 0   \vspace{5pt}\\
0 & 0 & 0            & 0
\end{array}\right), \label{eqappa25}\\
J_L&=&\left( \begin{array}{cccc}
0 & 0            & 0 & 0 \vspace{5pt}\\
0 & 0            & 0 & 0 \vspace{5pt}\\
0 & t_3+\gamma/2 & 0 & 0 \vspace{5pt}\\
t_2 & 0          & 0 & 0
\end{array}\right), \label{eqappa26}\\
M&=&\left( \begin{array}{cccc}
0            & t_1 & 0   & t_3+\gamma/2 \vspace{5pt}\\
t_1          & 0   & t_2 & 0            \vspace{5pt}\\
0            & t_2 & 0   & t_1          \vspace{5pt}\\
t_3-\gamma/2 & 0   & t_1 & 0
\end{array}\right). \label{eqappa27}
\end{eqnarray}
We note that the rank of both the hopping matrices is $2$, which means that the size of the transfer matrix becomes $4$. Let $\rho_r~\left(r=1,\dots,4\right)$ denote the eigenvalues of the transfer matrix. The condition for the generalized Brillouin zone can be obtained as follows:
\begin{equation}
\left|\rho_2\right|=\left|\rho_3\right|,
\label{eqappa28}
\end{equation}
where $\left|\rho_1\right|\leq\cdots\leq\left|\rho_4\right|$ is satisfied.

We calculate the generalized Brillouin zone [Fig.~\ref{figappa1}(c)] and the continuum energy bands [Fig.~\ref{figappa1}(d)] based on Eq.~(\ref{eqappa28}). We indeed confirm that the continuum energy bands reproduces the eigenenergies of a finite-size system with open boundary conditions.

%
%

%

\providecommand{\noopsort}[1]{}\providecommand{\singleletter}[1]{#1}%

\begin{thebibliography}{49}%
\makeatletter
\providecommand \@ifxundefined [1]{%
 \@ifx{#1\undefined}
}%
\providecommand \@ifnum [1]{%
 \ifnum #1\expandafter \@firstoftwo
 \else \expandafter \@secondoftwo
 \fi
}%
\providecommand \@ifx [1]{%
 \ifx #1\expandafter \@firstoftwo
 \else \expandafter \@secondoftwo
 \fi
}%
\providecommand \natexlab [1]{#1}%
\providecommand \enquote  [1]{``#1''}%
\providecommand \bibnamefont  [1]{#1}%
\providecommand \bibfnamefont [1]{#1}%
\providecommand \citenamefont [1]{#1}%
\providecommand \href@noop [0]{\@secondoftwo}%
\providecommand \href [0]{\begingroup \@sanitize@url \@href}%
\providecommand \@href[1]{\@@startlink{#1}\@@href}%
\providecommand \@@href[1]{\endgroup#1\@@endlink}%
\providecommand \@sanitize@url [0]{\catcode `\\12\catcode `\$12\catcode
  `\&12\catcode `\#12\catcode `\^12\catcode `\_12\catcode `\%12\relax}%
\providecommand \@@startlink[1]{}%
\providecommand \@@endlink[0]{}%
\providecommand \url  [0]{\begingroup\@sanitize@url \@url }%
\providecommand \@url [1]{\endgroup\@href {#1}{\urlprefix }}%
\providecommand \urlprefix  [0]{URL }%
\providecommand \Eprint [0]{\href }%
\providecommand \doibase [0]{https://doi.org/}%
\providecommand \selectlanguage [0]{\@gobble}%
\providecommand \bibinfo  [0]{\@secondoftwo}%
\providecommand \bibfield  [0]{\@secondoftwo}%
\providecommand \translation [1]{[#1]}%
\providecommand \BibitemOpen [0]{}%
\providecommand \bibitemStop [0]{}%
\providecommand \bibitemNoStop [0]{.\EOS\space}%
\providecommand \EOS [0]{\spacefactor3000\relax}%
\providecommand \BibitemShut  [1]{\csname bibitem#1\endcsname}%
\let\auto@bib@innerbib\@empty
\bibitem [{\citenamefont {Ashida}\ \emph {et~al.}(2020)\citenamefont {Ashida},
  \citenamefont {Gong},\ and\ \citenamefont {Ueda}}]{Ashida2020}%
  \BibitemOpen
  \bibfield  {author} {\bibinfo {author} {\bibfnamefont {Y.}~\bibnamefont
  {Ashida}}, \bibinfo {author} {\bibfnamefont {Z.}~\bibnamefont {Gong}},\ and\
  \bibinfo {author} {\bibfnamefont {M.}~\bibnamefont {Ueda}},\ }\href@noop {}
  {\bibfield  {journal} {\bibinfo  {journal} {Adv. Phys.}\ }\textbf {\bibinfo
  {volume} {69}},\ \bibinfo {pages} {249} (\bibinfo {year} {2020})}\BibitemShut
  {NoStop}%
\bibitem [{\citenamefont {Bergholtz}\ \emph {et~al.}(2021)\citenamefont
  {Bergholtz}, \citenamefont {Budich},\ and\ \citenamefont
  {Kunst}}]{Bergholtz2021}%
  \BibitemOpen
  \bibfield  {author} {\bibinfo {author} {\bibfnamefont {E.~J.}\ \bibnamefont
  {Bergholtz}}, \bibinfo {author} {\bibfnamefont {J.~C.}\ \bibnamefont
  {Budich}},\ and\ \bibinfo {author} {\bibfnamefont {F.~K.}\ \bibnamefont
  {Kunst}},\ }\href {https://doi.org/10.1103/RevModPhys.93.015005} {\bibfield
  {journal} {\bibinfo  {journal} {Rev. Mod. Phys.}\ }\textbf {\bibinfo {volume}
  {93}},\ \bibinfo {pages} {015005} (\bibinfo {year} {2021})}\BibitemShut
  {NoStop}%
\bibitem [{\citenamefont {Martinez~Alvarez}\ \emph {et~al.}(2018)\citenamefont
  {Martinez~Alvarez}, \citenamefont {Barrios~Vargas},\ and\ \citenamefont
  {Foa~Torres}}]{Martinez2018}%
  \BibitemOpen
  \bibfield  {author} {\bibinfo {author} {\bibfnamefont {V.~M.}\ \bibnamefont
  {Martinez~Alvarez}}, \bibinfo {author} {\bibfnamefont {J.~E.}\ \bibnamefont
  {Barrios~Vargas}},\ and\ \bibinfo {author} {\bibfnamefont {L.~E.~F.}\
  \bibnamefont {Foa~Torres}},\ }\href
  {https://doi.org/10.1103/PhysRevB.97.121401} {\bibfield  {journal} {\bibinfo
  {journal} {Phys. Rev. B}\ }\textbf {\bibinfo {volume} {97}},\ \bibinfo
  {pages} {121401(R)} (\bibinfo {year} {2018})}\BibitemShut {NoStop}%
\bibitem [{\citenamefont {Kunst}\ \emph {et~al.}(2018)\citenamefont {Kunst},
  \citenamefont {Edvardsson}, \citenamefont {Budich},\ and\ \citenamefont
  {Bergholtz}}]{Kunst2018}%
  \BibitemOpen
  \bibfield  {author} {\bibinfo {author} {\bibfnamefont {F.~K.}\ \bibnamefont
  {Kunst}}, \bibinfo {author} {\bibfnamefont {E.}~\bibnamefont {Edvardsson}},
  \bibinfo {author} {\bibfnamefont {J.~C.}\ \bibnamefont {Budich}},\ and\
  \bibinfo {author} {\bibfnamefont {E.~J.}\ \bibnamefont {Bergholtz}},\ }\href
  {https://doi.org/10.1103/PhysRevLett.121.026808} {\bibfield  {journal}
  {\bibinfo  {journal} {Phys. Rev. Lett.}\ }\textbf {\bibinfo {volume} {121}},\
  \bibinfo {pages} {026808} (\bibinfo {year} {2018})}\BibitemShut {NoStop}%
\bibitem [{\citenamefont {Yao}\ and\ \citenamefont {Wang}(2018)}]{Yao2018}%
  \BibitemOpen
  \bibfield  {author} {\bibinfo {author} {\bibfnamefont {S.}~\bibnamefont
  {Yao}}\ and\ \bibinfo {author} {\bibfnamefont {Z.}~\bibnamefont {Wang}},\
  }\href {https://doi.org/10.1103/PhysRevLett.121.086803} {\bibfield  {journal}
  {\bibinfo  {journal} {Phys. Rev. Lett.}\ }\textbf {\bibinfo {volume} {121}},\
  \bibinfo {pages} {086803} (\bibinfo {year} {2018})}\BibitemShut {NoStop}%
\bibitem [{\citenamefont {Song}\ \emph {et~al.}(2019)\citenamefont {Song},
  \citenamefont {Yao},\ and\ \citenamefont {Wang}}]{Song2019}%
  \BibitemOpen
  \bibfield  {author} {\bibinfo {author} {\bibfnamefont {F.}~\bibnamefont
  {Song}}, \bibinfo {author} {\bibfnamefont {S.}~\bibnamefont {Yao}},\ and\
  \bibinfo {author} {\bibfnamefont {Z.}~\bibnamefont {Wang}},\ }\href
  {https://doi.org/10.1103/PhysRevLett.123.170401} {\bibfield  {journal}
  {\bibinfo  {journal} {Phys. Rev. Lett.}\ }\textbf {\bibinfo {volume} {123}},\
  \bibinfo {pages} {170401} (\bibinfo {year} {2019})}\BibitemShut {NoStop}%
\bibitem [{\citenamefont {Borgnia}\ \emph {et~al.}(2020)\citenamefont
  {Borgnia}, \citenamefont {Kruchkov},\ and\ \citenamefont
  {Slager}}]{Borgnia2020}%
  \BibitemOpen
  \bibfield  {author} {\bibinfo {author} {\bibfnamefont {D.~S.}\ \bibnamefont
  {Borgnia}}, \bibinfo {author} {\bibfnamefont {A.~J.}\ \bibnamefont
  {Kruchkov}},\ and\ \bibinfo {author} {\bibfnamefont {R.-J.}\ \bibnamefont
  {Slager}},\ }\href {https://doi.org/10.1103/PhysRevLett.124.056802}
  {\bibfield  {journal} {\bibinfo  {journal} {Phys. Rev. Lett.}\ }\textbf
  {\bibinfo {volume} {124}},\ \bibinfo {pages} {056802} (\bibinfo {year}
  {2020})}\BibitemShut {NoStop}%
\bibitem [{\citenamefont {Okuma}\ \emph {et~al.}(2020)\citenamefont {Okuma},
  \citenamefont {Kawabata}, \citenamefont {Shiozaki},\ and\ \citenamefont
  {Sato}}]{Okuma2020}%
  \BibitemOpen
  \bibfield  {author} {\bibinfo {author} {\bibfnamefont {N.}~\bibnamefont
  {Okuma}}, \bibinfo {author} {\bibfnamefont {K.}~\bibnamefont {Kawabata}},
  \bibinfo {author} {\bibfnamefont {K.}~\bibnamefont {Shiozaki}},\ and\
  \bibinfo {author} {\bibfnamefont {M.}~\bibnamefont {Sato}},\ }\href
  {https://doi.org/10.1103/PhysRevLett.124.086801} {\bibfield  {journal}
  {\bibinfo  {journal} {Phys. Rev. Lett.}\ }\textbf {\bibinfo {volume} {124}},\
  \bibinfo {pages} {086801} (\bibinfo {year} {2020})}\BibitemShut {NoStop}%
\bibitem [{\citenamefont {Zhang}\ \emph {et~al.}(2020)\citenamefont {Zhang},
  \citenamefont {Yang},\ and\ \citenamefont {Fang}}]{Zhang2020}%
  \BibitemOpen
  \bibfield  {author} {\bibinfo {author} {\bibfnamefont {K.}~\bibnamefont
  {Zhang}}, \bibinfo {author} {\bibfnamefont {Z.}~\bibnamefont {Yang}},\ and\
  \bibinfo {author} {\bibfnamefont {C.}~\bibnamefont {Fang}},\ }\href
  {https://doi.org/10.1103/PhysRevLett.125.126402} {\bibfield  {journal}
  {\bibinfo  {journal} {Phys. Rev. Lett.}\ }\textbf {\bibinfo {volume} {125}},\
  \bibinfo {pages} {126402} (\bibinfo {year} {2020})}\BibitemShut {NoStop}%
\bibitem [{\citenamefont {Yi}\ and\ \citenamefont {Yang}(2020)}]{Yi2020}%
  \BibitemOpen
  \bibfield  {author} {\bibinfo {author} {\bibfnamefont {Y.}~\bibnamefont
  {Yi}}\ and\ \bibinfo {author} {\bibfnamefont {Z.}~\bibnamefont {Yang}},\
  }\href {https://doi.org/10.1103/PhysRevLett.125.186802} {\bibfield  {journal}
  {\bibinfo  {journal} {Phys. Rev. Lett.}\ }\textbf {\bibinfo {volume} {125}},\
  \bibinfo {pages} {186802} (\bibinfo {year} {2020})}\BibitemShut {NoStop}%
\bibitem [{\citenamefont {Brandenbourger}\ \emph {et~al.}(2019)\citenamefont
  {Brandenbourger}, \citenamefont {Locsin}, \citenamefont {Lerner},\ and\
  \citenamefont {Coulais}}]{Brandenbourger2019}%
  \BibitemOpen
  \bibfield  {author} {\bibinfo {author} {\bibfnamefont {M.}~\bibnamefont
  {Brandenbourger}}, \bibinfo {author} {\bibfnamefont {X.}~\bibnamefont
  {Locsin}}, \bibinfo {author} {\bibfnamefont {E.}~\bibnamefont {Lerner}},\
  and\ \bibinfo {author} {\bibfnamefont {C.}~\bibnamefont {Coulais}},\
  }\href@noop {} {\bibfield  {journal} {\bibinfo  {journal} {Nat. Commun.}\
  }\textbf {\bibinfo {volume} {10}},\ \bibinfo {pages} {4608} (\bibinfo {year}
  {2019})}\BibitemShut {NoStop}%
\bibitem [{\citenamefont {Xiao}\ \emph {et~al.}(2020)\citenamefont {Xiao},
  \citenamefont {Deng}, \citenamefont {Wang}, \citenamefont {Zhu},
  \citenamefont {Wang}, \citenamefont {Yi},\ and\ \citenamefont
  {Xue}}]{Xiao2020}%
  \BibitemOpen
  \bibfield  {author} {\bibinfo {author} {\bibfnamefont {L.}~\bibnamefont
  {Xiao}}, \bibinfo {author} {\bibfnamefont {T.}~\bibnamefont {Deng}}, \bibinfo
  {author} {\bibfnamefont {K.}~\bibnamefont {Wang}}, \bibinfo {author}
  {\bibfnamefont {G.}~\bibnamefont {Zhu}}, \bibinfo {author} {\bibfnamefont
  {Z.}~\bibnamefont {Wang}}, \bibinfo {author} {\bibfnamefont {W.}~\bibnamefont
  {Yi}},\ and\ \bibinfo {author} {\bibfnamefont {P.}~\bibnamefont {Xue}},\
  }\href@noop {} {\bibfield  {journal} {\bibinfo  {journal} {Nat. Phys.}\
  }\textbf {\bibinfo {volume} {16}},\ \bibinfo {pages} {761} (\bibinfo {year}
  {2020})}\BibitemShut {NoStop}%
\bibitem [{\citenamefont {Weidemann}\ \emph {et~al.}(2020)\citenamefont
  {Weidemann}, \citenamefont {Kremer}, \citenamefont {Helbig}, \citenamefont
  {Hofmann}, \citenamefont {Stegmaier}, \citenamefont {Greiter}, \citenamefont
  {Thomale},\ and\ \citenamefont {Szameit}}]{Weidemann2020}%
  \BibitemOpen
  \bibfield  {author} {\bibinfo {author} {\bibfnamefont {S.}~\bibnamefont
  {Weidemann}}, \bibinfo {author} {\bibfnamefont {M.}~\bibnamefont {Kremer}},
  \bibinfo {author} {\bibfnamefont {T.}~\bibnamefont {Helbig}}, \bibinfo
  {author} {\bibfnamefont {T.}~\bibnamefont {Hofmann}}, \bibinfo {author}
  {\bibfnamefont {A.}~\bibnamefont {Stegmaier}}, \bibinfo {author}
  {\bibfnamefont {M.}~\bibnamefont {Greiter}}, \bibinfo {author} {\bibfnamefont
  {R.}~\bibnamefont {Thomale}},\ and\ \bibinfo {author} {\bibfnamefont
  {A.}~\bibnamefont {Szameit}},\ }\href@noop {} {\bibfield  {journal} {\bibinfo
   {journal} {Science}\ }\textbf {\bibinfo {volume} {368}},\ \bibinfo {pages}
  {311} (\bibinfo {year} {2020})}\BibitemShut {NoStop}%
\bibitem [{\citenamefont {Helbig}\ \emph {et~al.}(2020)\citenamefont {Helbig},
  \citenamefont {Hofmann}, \citenamefont {Imhof}, \citenamefont {Abdelghany},
  \citenamefont {Kiessling}, \citenamefont {Molenkamp}, \citenamefont {Lee},
  \citenamefont {Szameit}, \citenamefont {Greiter},\ and\ \citenamefont
  {Thomale}}]{Helbig2020}%
  \BibitemOpen
  \bibfield  {author} {\bibinfo {author} {\bibfnamefont {T.}~\bibnamefont
  {Helbig}}, \bibinfo {author} {\bibfnamefont {T.}~\bibnamefont {Hofmann}},
  \bibinfo {author} {\bibfnamefont {S.}~\bibnamefont {Imhof}}, \bibinfo
  {author} {\bibfnamefont {M.}~\bibnamefont {Abdelghany}}, \bibinfo {author}
  {\bibfnamefont {T.}~\bibnamefont {Kiessling}}, \bibinfo {author}
  {\bibfnamefont {L.}~\bibnamefont {Molenkamp}}, \bibinfo {author}
  {\bibfnamefont {C.}~\bibnamefont {Lee}}, \bibinfo {author} {\bibfnamefont
  {A.}~\bibnamefont {Szameit}}, \bibinfo {author} {\bibfnamefont
  {M.}~\bibnamefont {Greiter}},\ and\ \bibinfo {author} {\bibfnamefont
  {R.}~\bibnamefont {Thomale}},\ }\href@noop {} {\bibfield  {journal} {\bibinfo
   {journal} {Nat. Phys.}\ }\textbf {\bibinfo {volume} {16}},\ \bibinfo {pages}
  {747} (\bibinfo {year} {2020})}\BibitemShut {NoStop}%
\bibitem [{\citenamefont {Hofmann}\ \emph {et~al.}(2020)\citenamefont
  {Hofmann}, \citenamefont {Helbig}, \citenamefont {Schindler}, \citenamefont
  {Salgo}, \citenamefont {Brzezi\ifmmode~\acute{n}\else \'{n}\fi{}ska},
  \citenamefont {Greiter}, \citenamefont {Kiessling}, \citenamefont {Wolf},
  \citenamefont {Vollhardt}, \citenamefont {Kaba\ifmmode~\check{s}\else
  \v{s}\fi{}i}, \citenamefont {Lee}, \citenamefont {Bilu\ifmmode \check{s}\else
  \v{s}\fi{}i\ifmmode~\acute{c}\else \'{c}\fi{}}, \citenamefont {Thomale},\
  and\ \citenamefont {Neupert}}]{Hofmann2020}%
  \BibitemOpen
  \bibfield  {author} {\bibinfo {author} {\bibfnamefont {T.}~\bibnamefont
  {Hofmann}}, \bibinfo {author} {\bibfnamefont {T.}~\bibnamefont {Helbig}},
  \bibinfo {author} {\bibfnamefont {F.}~\bibnamefont {Schindler}}, \bibinfo
  {author} {\bibfnamefont {N.}~\bibnamefont {Salgo}}, \bibinfo {author}
  {\bibfnamefont {M.}~\bibnamefont {Brzezi\ifmmode~\acute{n}\else
  \'{n}\fi{}ska}}, \bibinfo {author} {\bibfnamefont {M.}~\bibnamefont
  {Greiter}}, \bibinfo {author} {\bibfnamefont {T.}~\bibnamefont {Kiessling}},
  \bibinfo {author} {\bibfnamefont {D.}~\bibnamefont {Wolf}}, \bibinfo {author}
  {\bibfnamefont {A.}~\bibnamefont {Vollhardt}}, \bibinfo {author}
  {\bibfnamefont {A.}~\bibnamefont {Kaba\ifmmode~\check{s}\else \v{s}\fi{}i}},
  \bibinfo {author} {\bibfnamefont {C.~H.}\ \bibnamefont {Lee}}, \bibinfo
  {author} {\bibfnamefont {A.}~\bibnamefont {Bilu\ifmmode \check{s}\else
  \v{s}\fi{}i\ifmmode~\acute{c}\else \'{c}\fi{}}}, \bibinfo {author}
  {\bibfnamefont {R.}~\bibnamefont {Thomale}},\ and\ \bibinfo {author}
  {\bibfnamefont {T.}~\bibnamefont {Neupert}},\ }\href
  {https://doi.org/10.1103/PhysRevResearch.2.023265} {\bibfield  {journal}
  {\bibinfo  {journal} {Phys. Rev. Research}\ }\textbf {\bibinfo {volume}
  {2}},\ \bibinfo {pages} {023265} (\bibinfo {year} {2020})}\BibitemShut
  {NoStop}%
\bibitem [{\citenamefont {Ghatak}\ \emph {et~al.}(2020)\citenamefont {Ghatak},
  \citenamefont {Brandenbourger}, \citenamefont {van Wezel},\ and\
  \citenamefont {Coulais}}]{Ghatak2020}%
  \BibitemOpen
  \bibfield  {author} {\bibinfo {author} {\bibfnamefont {A.}~\bibnamefont
  {Ghatak}}, \bibinfo {author} {\bibfnamefont {M.}~\bibnamefont
  {Brandenbourger}}, \bibinfo {author} {\bibfnamefont {J.}~\bibnamefont {van
  Wezel}},\ and\ \bibinfo {author} {\bibfnamefont {C.}~\bibnamefont
  {Coulais}},\ }\href@noop {} {\bibfield  {journal} {\bibinfo  {journal} {Proc.
  Nat. Ac. Sc. USA}\ }\textbf {\bibinfo {volume} {117}},\ \bibinfo {pages}
  {29561} (\bibinfo {year} {2020})}\BibitemShut {NoStop}%
\bibitem [{\citenamefont {Chen}\ \emph {et~al.}(2021)\citenamefont {Chen},
  \citenamefont {Li}, \citenamefont {Scheibner}, \citenamefont {Vitelli},\ and\
  \citenamefont {Huang}}]{Chen2021}%
  \BibitemOpen
  \bibfield  {author} {\bibinfo {author} {\bibfnamefont {Y.}~\bibnamefont
  {Chen}}, \bibinfo {author} {\bibfnamefont {X.}~\bibnamefont {Li}}, \bibinfo
  {author} {\bibfnamefont {C.}~\bibnamefont {Scheibner}}, \bibinfo {author}
  {\bibfnamefont {V.}~\bibnamefont {Vitelli}},\ and\ \bibinfo {author}
  {\bibfnamefont {G.}~\bibnamefont {Huang}},\ }\href@noop {} {\bibfield
  {journal} {\bibinfo  {journal} {Nat. Commun.}\ }\textbf {\bibinfo {volume}
  {12}},\ \bibinfo {pages} {5935} (\bibinfo {year} {2021})}\BibitemShut
  {NoStop}%
\bibitem [{\citenamefont {Zhang}\ \emph {et~al.}(2021)\citenamefont {Zhang},
  \citenamefont {Yang}, \citenamefont {Ge}, \citenamefont {Guan}, \citenamefont
  {Chen}, \citenamefont {Yan}, \citenamefont {Chen}, \citenamefont {Xi},
  \citenamefont {Li}, \citenamefont {Jia}, \citenamefont {Yuan}, \citenamefont
  {Chen},\ and\ \citenamefont {Zhang}}]{Zhang2021}%
  \BibitemOpen
  \bibfield  {author} {\bibinfo {author} {\bibfnamefont {L.}~\bibnamefont
  {Zhang}}, \bibinfo {author} {\bibfnamefont {Y.}~\bibnamefont {Yang}},
  \bibinfo {author} {\bibfnamefont {Y.}~\bibnamefont {Ge}}, \bibinfo {author}
  {\bibfnamefont {Y.-J.}\ \bibnamefont {Guan}}, \bibinfo {author}
  {\bibfnamefont {Q.}~\bibnamefont {Chen}}, \bibinfo {author} {\bibfnamefont
  {Q.}~\bibnamefont {Yan}}, \bibinfo {author} {\bibfnamefont {F.}~\bibnamefont
  {Chen}}, \bibinfo {author} {\bibfnamefont {R.}~\bibnamefont {Xi}}, \bibinfo
  {author} {\bibfnamefont {Y.}~\bibnamefont {Li}}, \bibinfo {author}
  {\bibfnamefont {D.}~\bibnamefont {Jia}}, \bibinfo {author} {\bibfnamefont
  {S.-Q.}\ \bibnamefont {Yuan}}, \bibinfo {author} {\bibfnamefont
  {H.}~\bibnamefont {Chen}},\ and\ \bibinfo {author} {\bibfnamefont
  {B.}~\bibnamefont {Zhang}},\ }\href@noop {} {\bibfield  {journal} {\bibinfo
  {journal} {Nat. Commun.}\ }\textbf {\bibinfo {volume} {12}},\ \bibinfo
  {pages} {6297} (\bibinfo {year} {2021})}\BibitemShut {NoStop}%
\bibitem [{\citenamefont {Wang}\ \emph {et~al.}(2022)\citenamefont {Wang},
  \citenamefont {Wang},\ and\ \citenamefont {Ma}}]{Wang2022}%
  \BibitemOpen
  \bibfield  {author} {\bibinfo {author} {\bibfnamefont {W.}~\bibnamefont
  {Wang}}, \bibinfo {author} {\bibfnamefont {X.}~\bibnamefont {Wang}},\ and\
  \bibinfo {author} {\bibfnamefont {G.}~\bibnamefont {Ma}},\ }\href@noop {}
  {\bibfield  {journal} {\bibinfo  {journal} {Nature}\ }\textbf {\bibinfo
  {volume} {608}},\ \bibinfo {pages} {50} (\bibinfo {year} {2022})}\BibitemShut
  {NoStop}%
\bibitem [{\citenamefont {Liang}\ \emph {et~al.}(2022)\citenamefont {Liang},
  \citenamefont {Xie}, \citenamefont {Dong}, \citenamefont {Li}, \citenamefont
  {Li}, \citenamefont {Gadway}, \citenamefont {Yi},\ and\ \citenamefont
  {Yan}}]{Liang2022}%
  \BibitemOpen
  \bibfield  {author} {\bibinfo {author} {\bibfnamefont {Q.}~\bibnamefont
  {Liang}}, \bibinfo {author} {\bibfnamefont {D.}~\bibnamefont {Xie}}, \bibinfo
  {author} {\bibfnamefont {Z.}~\bibnamefont {Dong}}, \bibinfo {author}
  {\bibfnamefont {H.}~\bibnamefont {Li}}, \bibinfo {author} {\bibfnamefont
  {H.}~\bibnamefont {Li}}, \bibinfo {author} {\bibfnamefont {B.}~\bibnamefont
  {Gadway}}, \bibinfo {author} {\bibfnamefont {W.}~\bibnamefont {Yi}},\ and\
  \bibinfo {author} {\bibfnamefont {B.}~\bibnamefont {Yan}},\ }\href
  {https://doi.org/10.1103/PhysRevLett.129.070401} {\bibfield  {journal}
  {\bibinfo  {journal} {Phys. Rev. Lett.}\ }\textbf {\bibinfo {volume} {129}},\
  \bibinfo {pages} {070401} (\bibinfo {year} {2022})}\BibitemShut {NoStop}%
\bibitem [{\citenamefont {Gu}\ \emph {et~al.}(2022)\citenamefont {Gu},
  \citenamefont {Gao}, \citenamefont {Xue}, \citenamefont {Li}, \citenamefont
  {Su},\ and\ \citenamefont {Zhu}}]{Gu2022}%
  \BibitemOpen
  \bibfield  {author} {\bibinfo {author} {\bibfnamefont {Z.}~\bibnamefont
  {Gu}}, \bibinfo {author} {\bibfnamefont {H.}~\bibnamefont {Gao}}, \bibinfo
  {author} {\bibfnamefont {H.}~\bibnamefont {Xue}}, \bibinfo {author}
  {\bibfnamefont {J.}~\bibnamefont {Li}}, \bibinfo {author} {\bibfnamefont
  {Z.}~\bibnamefont {Su}},\ and\ \bibinfo {author} {\bibfnamefont
  {J.}~\bibnamefont {Zhu}},\ }\href@noop {} {\bibfield  {journal} {\bibinfo
  {journal} {Nat. Commun.}\ }\textbf {\bibinfo {volume} {13}},\ \bibinfo
  {pages} {7668} (\bibinfo {year} {2022})}\BibitemShut {NoStop}%
\bibitem [{\citenamefont {Yokomizo}\ and\ \citenamefont
  {Murakami}(2019)}]{Yokomizo2019}%
  \BibitemOpen
  \bibfield  {author} {\bibinfo {author} {\bibfnamefont {K.}~\bibnamefont
  {Yokomizo}}\ and\ \bibinfo {author} {\bibfnamefont {S.}~\bibnamefont
  {Murakami}},\ }\href {https://doi.org/10.1103/PhysRevLett.123.066404}
  {\bibfield  {journal} {\bibinfo  {journal} {Phys. Rev. Lett.}\ }\textbf
  {\bibinfo {volume} {123}},\ \bibinfo {pages} {066404} (\bibinfo {year}
  {2019})}\BibitemShut {NoStop}%
\bibitem [{\citenamefont {Yokomizo}\ and\ \citenamefont
  {Murakami}(2020)}]{Yokomizo2020}%
  \BibitemOpen
  \bibfield  {author} {\bibinfo {author} {\bibfnamefont {K.}~\bibnamefont
  {Yokomizo}}\ and\ \bibinfo {author} {\bibfnamefont {S.}~\bibnamefont
  {Murakami}},\ }\href@noop {} {\bibfield  {journal} {\bibinfo  {journal}
  {Prog. Theor. Exp. Phys.}\ }\textbf {\bibinfo {volume} {2020}},\ \bibinfo
  {pages} {12A102} (\bibinfo {year} {2020})}\BibitemShut {NoStop}%
\bibitem [{\citenamefont {Kawabata}\ \emph {et~al.}(2020)\citenamefont
  {Kawabata}, \citenamefont {Okuma},\ and\ \citenamefont
  {Sato}}]{Kawabata2020}%
  \BibitemOpen
  \bibfield  {author} {\bibinfo {author} {\bibfnamefont {K.}~\bibnamefont
  {Kawabata}}, \bibinfo {author} {\bibfnamefont {N.}~\bibnamefont {Okuma}},\
  and\ \bibinfo {author} {\bibfnamefont {M.}~\bibnamefont {Sato}},\ }\href
  {https://doi.org/10.1103/PhysRevB.101.195147} {\bibfield  {journal} {\bibinfo
   {journal} {Phys. Rev. B}\ }\textbf {\bibinfo {volume} {101}},\ \bibinfo
  {pages} {195147} (\bibinfo {year} {2020})}\BibitemShut {NoStop}%
\bibitem [{\citenamefont {Yang}\ \emph {et~al.}(2020)\citenamefont {Yang},
  \citenamefont {Zhang}, \citenamefont {Fang},\ and\ \citenamefont
  {Hu}}]{Yang2020}%
  \BibitemOpen
  \bibfield  {author} {\bibinfo {author} {\bibfnamefont {Z.}~\bibnamefont
  {Yang}}, \bibinfo {author} {\bibfnamefont {K.}~\bibnamefont {Zhang}},
  \bibinfo {author} {\bibfnamefont {C.}~\bibnamefont {Fang}},\ and\ \bibinfo
  {author} {\bibfnamefont {J.}~\bibnamefont {Hu}},\ }\href
  {https://doi.org/10.1103/PhysRevLett.125.226402} {\bibfield  {journal}
  {\bibinfo  {journal} {Phys. Rev. Lett.}\ }\textbf {\bibinfo {volume} {125}},\
  \bibinfo {pages} {226402} (\bibinfo {year} {2020})}\BibitemShut {NoStop}%
\bibitem [{\citenamefont {Yokomizo}\ \emph {et~al.}(2022)\citenamefont
  {Yokomizo}, \citenamefont {Yoda},\ and\ \citenamefont
  {Murakami}}]{Yokomizo2022}%
  \BibitemOpen
  \bibfield  {author} {\bibinfo {author} {\bibfnamefont {K.}~\bibnamefont
  {Yokomizo}}, \bibinfo {author} {\bibfnamefont {T.}~\bibnamefont {Yoda}},\
  and\ \bibinfo {author} {\bibfnamefont {S.}~\bibnamefont {Murakami}},\ }\href
  {https://doi.org/10.1103/PhysRevResearch.4.023089} {\bibfield  {journal}
  {\bibinfo  {journal} {Phys. Rev. Research}\ }\textbf {\bibinfo {volume}
  {4}},\ \bibinfo {pages} {023089} (\bibinfo {year} {2022})}\BibitemShut
  {NoStop}%
\bibitem [{\citenamefont {Hu}\ \emph {et~al.}()\citenamefont {Hu},
  \citenamefont {Huang}, \citenamefont {Xue},\ and\ \citenamefont
  {Wang}}]{Hu2023}%
  \BibitemOpen
  \bibfield  {author} {\bibinfo {author} {\bibfnamefont {Y.-M.}\ \bibnamefont
  {Hu}}, \bibinfo {author} {\bibfnamefont {Y.-Q.}\ \bibnamefont {Huang}},
  \bibinfo {author} {\bibfnamefont {W.-T.}\ \bibnamefont {Xue}},\ and\ \bibinfo
  {author} {\bibfnamefont {Z.}~\bibnamefont {Wang}},\ }\href@noop {} {\bibinfo
  {journal} {arXiv:2310.08572}\ }\BibitemShut {NoStop}%
\bibitem [{\citenamefont {Inoue}\ and\ \citenamefont
  {Murakami}(2019)}]{Inoue2019}%
  \BibitemOpen
\bibfield  {journal} {  }\bibfield  {author} {\bibinfo {author} {\bibfnamefont
  {T.}~\bibnamefont {Inoue}}\ and\ \bibinfo {author} {\bibfnamefont
  {S.}~\bibnamefont {Murakami}},\ }\href
  {https://doi.org/10.1103/PhysRevB.99.195443} {\bibfield  {journal} {\bibinfo
  {journal} {Phys. Rev. B}\ }\textbf {\bibinfo {volume} {99}},\ \bibinfo
  {pages} {195443} (\bibinfo {year} {2019})}\BibitemShut {NoStop}%
\bibitem [{\citenamefont {Haldane}\ and\ \citenamefont
  {Raghu}(2008)}]{Haldane2008}%
  \BibitemOpen
  \bibfield  {author} {\bibinfo {author} {\bibfnamefont {F.~D.~M.}\
  \bibnamefont {Haldane}}\ and\ \bibinfo {author} {\bibfnamefont
  {S.}~\bibnamefont {Raghu}},\ }\href
  {https://doi.org/10.1103/PhysRevLett.100.013904} {\bibfield  {journal}
  {\bibinfo  {journal} {Phys. Rev. Lett.}\ }\textbf {\bibinfo {volume} {100}},\
  \bibinfo {pages} {013904} (\bibinfo {year} {2008})}\BibitemShut {NoStop}%
\bibitem [{\citenamefont {Raman}\ and\ \citenamefont {Fan}(2010)}]{Raman2010}%
  \BibitemOpen
  \bibfield  {author} {\bibinfo {author} {\bibfnamefont {A.}~\bibnamefont
  {Raman}}\ and\ \bibinfo {author} {\bibfnamefont {S.}~\bibnamefont {Fan}},\
  }\href {https://doi.org/10.1103/PhysRevLett.104.087401} {\bibfield  {journal}
  {\bibinfo  {journal} {Phys. Rev. Lett.}\ }\textbf {\bibinfo {volume} {104}},\
  \bibinfo {pages} {087401} (\bibinfo {year} {2010})}\BibitemShut {NoStop}%
\bibitem [{\citenamefont {Shindou}\ \emph
  {et~al.}(2013{\natexlab{a}})\citenamefont {Shindou}, \citenamefont {Ohe},
  \citenamefont {Matsumoto}, \citenamefont {Murakami},\ and\ \citenamefont
  {Saitoh}}]{Shidou2013}%
  \BibitemOpen
  \bibfield  {author} {\bibinfo {author} {\bibfnamefont {R.}~\bibnamefont
  {Shindou}}, \bibinfo {author} {\bibfnamefont {J.-I.}\ \bibnamefont {Ohe}},
  \bibinfo {author} {\bibfnamefont {R.}~\bibnamefont {Matsumoto}}, \bibinfo
  {author} {\bibfnamefont {S.}~\bibnamefont {Murakami}},\ and\ \bibinfo
  {author} {\bibfnamefont {E.}~\bibnamefont {Saitoh}},\ }\href
  {https://doi.org/10.1103/PhysRevB.87.174402} {\bibfield  {journal} {\bibinfo
  {journal} {Phys. Rev. B}\ }\textbf {\bibinfo {volume} {87}},\ \bibinfo
  {pages} {174402} (\bibinfo {year} {2013}{\natexlab{a}})}\BibitemShut
  {NoStop}%
\bibitem [{\citenamefont {Shindou}\ \emph
  {et~al.}(2013{\natexlab{b}})\citenamefont {Shindou}, \citenamefont
  {Matsumoto}, \citenamefont {Murakami},\ and\ \citenamefont
  {Ohe}}]{Shidou2013v2}%
  \BibitemOpen
  \bibfield  {author} {\bibinfo {author} {\bibfnamefont {R.}~\bibnamefont
  {Shindou}}, \bibinfo {author} {\bibfnamefont {R.}~\bibnamefont {Matsumoto}},
  \bibinfo {author} {\bibfnamefont {S.}~\bibnamefont {Murakami}},\ and\
  \bibinfo {author} {\bibfnamefont {J.-i.}\ \bibnamefont {Ohe}},\ }\href
  {https://doi.org/10.1103/PhysRevB.87.174427} {\bibfield  {journal} {\bibinfo
  {journal} {Phys. Rev. B}\ }\textbf {\bibinfo {volume} {87}},\ \bibinfo
  {pages} {174427} (\bibinfo {year} {2013}{\natexlab{b}})}\BibitemShut
  {NoStop}%
\bibitem [{\citenamefont {Isobe}\ \emph {et~al.}()\citenamefont {Isobe},
  \citenamefont {Yoshida},\ and\ \citenamefont {Hatsugai}}]{Isobe2023v2}%
  \BibitemOpen
  \bibfield  {author} {\bibinfo {author} {\bibfnamefont {T.}~\bibnamefont
  {Isobe}}, \bibinfo {author} {\bibfnamefont {T.}~\bibnamefont {Yoshida}},\
  and\ \bibinfo {author} {\bibfnamefont {Y.}~\bibnamefont {Hatsugai}},\
  }\href@noop {} {\bibinfo  {journal} {arXiv:2310.12577}\ }\BibitemShut
  {NoStop}%
\bibitem [{\citenamefont {Isobe}\ \emph {et~al.}(2021)\citenamefont {Isobe},
  \citenamefont {Yoshida},\ and\ \citenamefont {Hatsugai}}]{Isobe2021}%
  \BibitemOpen
\bibfield  {journal} {  }\bibfield  {author} {\bibinfo {author} {\bibfnamefont
  {T.}~\bibnamefont {Isobe}}, \bibinfo {author} {\bibfnamefont
  {T.}~\bibnamefont {Yoshida}},\ and\ \bibinfo {author} {\bibfnamefont
  {Y.}~\bibnamefont {Hatsugai}},\ }\href
  {https://doi.org/10.1103/PhysRevB.104.L121105} {\bibfield  {journal}
  {\bibinfo  {journal} {Phys. Rev. B}\ }\textbf {\bibinfo {volume} {104}},\
  \bibinfo {pages} {L121105} (\bibinfo {year} {2021})}\BibitemShut {NoStop}%
\bibitem [{\citenamefont {Isobe}\ \emph {et~al.}(2023)\citenamefont {Isobe},
  \citenamefont {Yoshida},\ and\ \citenamefont {Hatsugai}}]{Isobe2023}%
  \BibitemOpen
  \bibfield  {author} {\bibinfo {author} {\bibfnamefont {T.}~\bibnamefont
  {Isobe}}, \bibinfo {author} {\bibfnamefont {T.}~\bibnamefont {Yoshida}},\
  and\ \bibinfo {author} {\bibfnamefont {Y.}~\bibnamefont {Hatsugai}},\
  }\href@noop {} {\bibfield  {journal} {\bibinfo  {journal} {Nanophotonics}\
  }\textbf {\bibinfo {volume} {12}},\ \bibinfo {pages} {2335} (\bibinfo {year}
  {2023})}\BibitemShut {NoStop}%
\bibitem [{\citenamefont {McDonald}\ \emph {et~al.}(2018)\citenamefont
  {McDonald}, \citenamefont {Pereg-Barnea},\ and\ \citenamefont
  {Clerk}}]{McDonald2018}%
  \BibitemOpen
  \bibfield  {author} {\bibinfo {author} {\bibfnamefont {A.}~\bibnamefont
  {McDonald}}, \bibinfo {author} {\bibfnamefont {T.}~\bibnamefont
  {Pereg-Barnea}},\ and\ \bibinfo {author} {\bibfnamefont {A.~A.}\ \bibnamefont
  {Clerk}},\ }\href {https://doi.org/10.1103/PhysRevX.8.041031} {\bibfield
  {journal} {\bibinfo  {journal} {Phys. Rev. X}\ }\textbf {\bibinfo {volume}
  {8}},\ \bibinfo {pages} {041031} (\bibinfo {year} {2018})}\BibitemShut
  {NoStop}%
\bibitem [{\citenamefont {Yokomizo}\ and\ \citenamefont
  {Murakami}(2021)}]{Yokomizo2021}%
  \BibitemOpen
  \bibfield  {author} {\bibinfo {author} {\bibfnamefont {K.}~\bibnamefont
  {Yokomizo}}\ and\ \bibinfo {author} {\bibfnamefont {S.}~\bibnamefont
  {Murakami}},\ }\href {https://doi.org/10.1103/PhysRevB.103.165123} {\bibfield
   {journal} {\bibinfo  {journal} {Phys. Rev. B}\ }\textbf {\bibinfo {volume}
  {103}},\ \bibinfo {pages} {165123} (\bibinfo {year} {2021})}\BibitemShut
  {NoStop}%
\bibitem [{\citenamefont {Rosa}\ and\ \citenamefont
  {Ruzzene}(2020)}]{Rosa2020}%
  \BibitemOpen
  \bibfield  {author} {\bibinfo {author} {\bibfnamefont {M.~I.}\ \bibnamefont
  {Rosa}}\ and\ \bibinfo {author} {\bibfnamefont {M.}~\bibnamefont {Ruzzene}},\
  }\href@noop {} {\bibfield  {journal} {\bibinfo  {journal} {New J. Phys.}\
  }\textbf {\bibinfo {volume} {22}},\ \bibinfo {pages} {053004} (\bibinfo
  {year} {2020})}\BibitemShut {NoStop}%
\bibitem [{\citenamefont {Braghini}\ \emph {et~al.}(2021)\citenamefont
  {Braghini}, \citenamefont {Villani}, \citenamefont {Rosa},\ and\
  \citenamefont {de~F~Arruda}}]{Braghini2021}%
  \BibitemOpen
  \bibfield  {author} {\bibinfo {author} {\bibfnamefont {D.}~\bibnamefont
  {Braghini}}, \bibinfo {author} {\bibfnamefont {L.~G.}\ \bibnamefont
  {Villani}}, \bibinfo {author} {\bibfnamefont {M.~I.}\ \bibnamefont {Rosa}},\
  and\ \bibinfo {author} {\bibfnamefont {J.~R.}\ \bibnamefont {de~F~Arruda}},\
  }\href@noop {} {\bibfield  {journal} {\bibinfo  {journal} {J. Phys. D}\
  }\textbf {\bibinfo {volume} {54}},\ \bibinfo {pages} {285302} (\bibinfo
  {year} {2021})}\BibitemShut {NoStop}%
\bibitem [{\citenamefont {Yan}\ \emph {et~al.}(2021)\citenamefont {Yan},
  \citenamefont {Chen},\ and\ \citenamefont {Yang}}]{Yan2021}%
  \BibitemOpen
  \bibfield  {author} {\bibinfo {author} {\bibfnamefont {Q.}~\bibnamefont
  {Yan}}, \bibinfo {author} {\bibfnamefont {H.}~\bibnamefont {Chen}},\ and\
  \bibinfo {author} {\bibfnamefont {Y.}~\bibnamefont {Yang}},\ }\href@noop {}
  {\bibfield  {journal} {\bibinfo  {journal} {Prog. Electromagn. Res.}\
  }\textbf {\bibinfo {volume} {172}},\ \bibinfo {pages} {33} (\bibinfo {year}
  {2021})}\BibitemShut {NoStop}%
\bibitem [{\citenamefont {Gear}\ \emph {et~al.}(2015)\citenamefont {Gear},
  \citenamefont {Liu}, \citenamefont {Chu}, \citenamefont {Rotter},\ and\
  \citenamefont {Li}}]{Gear2015}%
  \BibitemOpen
  \bibfield  {author} {\bibinfo {author} {\bibfnamefont {J.}~\bibnamefont
  {Gear}}, \bibinfo {author} {\bibfnamefont {F.}~\bibnamefont {Liu}}, \bibinfo
  {author} {\bibfnamefont {S.~T.}\ \bibnamefont {Chu}}, \bibinfo {author}
  {\bibfnamefont {S.}~\bibnamefont {Rotter}},\ and\ \bibinfo {author}
  {\bibfnamefont {J.}~\bibnamefont {Li}},\ }\href
  {https://doi.org/10.1103/PhysRevA.91.033825} {\bibfield  {journal} {\bibinfo
  {journal} {Phys. Rev. A}\ }\textbf {\bibinfo {volume} {91}},\ \bibinfo
  {pages} {033825} (\bibinfo {year} {2015})}\BibitemShut {NoStop}%
\bibitem [{\citenamefont {Droulias}\ \emph {et~al.}(2019)\citenamefont
  {Droulias}, \citenamefont {Katsantonis}, \citenamefont {Kafesaki},
  \citenamefont {Soukoulis},\ and\ \citenamefont {Economou}}]{Droulias2019}%
  \BibitemOpen
  \bibfield  {author} {\bibinfo {author} {\bibfnamefont {S.}~\bibnamefont
  {Droulias}}, \bibinfo {author} {\bibfnamefont {I.}~\bibnamefont
  {Katsantonis}}, \bibinfo {author} {\bibfnamefont {M.}~\bibnamefont
  {Kafesaki}}, \bibinfo {author} {\bibfnamefont {C.~M.}\ \bibnamefont
  {Soukoulis}},\ and\ \bibinfo {author} {\bibfnamefont {E.~N.}\ \bibnamefont
  {Economou}},\ }\href {https://doi.org/10.1103/PhysRevLett.122.213201}
  {\bibfield  {journal} {\bibinfo  {journal} {Phys. Rev. Lett.}\ }\textbf
  {\bibinfo {volume} {122}},\ \bibinfo {pages} {213201} (\bibinfo {year}
  {2019})}\BibitemShut {NoStop}%
\bibitem [{\citenamefont {Yoda}\ \emph {et~al.}()\citenamefont {Yoda},
  \citenamefont {Moritake}, \citenamefont {Takata}, \citenamefont {Yokomizo},
  \citenamefont {Murakami},\ and\ \citenamefont {Notomi}}]{Yoda2023}%
  \BibitemOpen
  \bibfield  {author} {\bibinfo {author} {\bibfnamefont {T.}~\bibnamefont
  {Yoda}}, \bibinfo {author} {\bibfnamefont {Y.}~\bibnamefont {Moritake}},
  \bibinfo {author} {\bibfnamefont {K.}~\bibnamefont {Takata}}, \bibinfo
  {author} {\bibfnamefont {K.}~\bibnamefont {Yokomizo}}, \bibinfo {author}
  {\bibfnamefont {S.}~\bibnamefont {Murakami}},\ and\ \bibinfo {author}
  {\bibfnamefont {M.}~\bibnamefont {Notomi}},\ }\href@noop {} {\bibinfo
  {journal} {arXiv:2303.05185}\ }\BibitemShut {NoStop}%
\bibitem [{\citenamefont {Dwivedi}\ and\ \citenamefont
  {Chua}(2016)}]{Dwivedi2016}%
  \BibitemOpen
\bibfield  {journal} {  }\bibfield  {author} {\bibinfo {author} {\bibfnamefont
  {V.}~\bibnamefont {Dwivedi}}\ and\ \bibinfo {author} {\bibfnamefont
  {V.}~\bibnamefont {Chua}},\ }\href
  {https://doi.org/10.1103/PhysRevB.93.134304} {\bibfield  {journal} {\bibinfo
  {journal} {Phys. Rev. B}\ }\textbf {\bibinfo {volume} {93}},\ \bibinfo
  {pages} {134304} (\bibinfo {year} {2016})}\BibitemShut {NoStop}%
\bibitem [{\citenamefont {Kunst}\ and\ \citenamefont
  {Dwivedi}(2019)}]{Kunst2019}%
  \BibitemOpen
  \bibfield  {author} {\bibinfo {author} {\bibfnamefont {F.~K.}\ \bibnamefont
  {Kunst}}\ and\ \bibinfo {author} {\bibfnamefont {V.}~\bibnamefont
  {Dwivedi}},\ }\href {https://doi.org/10.1103/PhysRevB.99.245116} {\bibfield
  {journal} {\bibinfo  {journal} {Phys. Rev. B}\ }\textbf {\bibinfo {volume}
  {99}},\ \bibinfo {pages} {245116} (\bibinfo {year} {2019})}\BibitemShut
  {NoStop}%
\bibitem [{\citenamefont {Wang}\ \emph {et~al.}(2019)\citenamefont {Wang},
  \citenamefont {Hou}, \citenamefont {Lu}, \citenamefont {Chen}, \citenamefont
  {Zhang},\ and\ \citenamefont {Chan}}]{Wang2019}%
  \BibitemOpen
  \bibfield  {author} {\bibinfo {author} {\bibfnamefont {S.}~\bibnamefont
  {Wang}}, \bibinfo {author} {\bibfnamefont {B.}~\bibnamefont {Hou}}, \bibinfo
  {author} {\bibfnamefont {W.}~\bibnamefont {Lu}}, \bibinfo {author}
  {\bibfnamefont {Y.}~\bibnamefont {Chen}}, \bibinfo {author} {\bibfnamefont
  {Z.}~\bibnamefont {Zhang}},\ and\ \bibinfo {author} {\bibfnamefont {C.~T.}\
  \bibnamefont {Chan}},\ }\href@noop {} {\bibfield  {journal} {\bibinfo
  {journal} {Nat. Commun.}\ }\textbf {\bibinfo {volume} {10}},\ \bibinfo
  {pages} {832} (\bibinfo {year} {2019})}\BibitemShut {NoStop}%
\bibitem [{\citenamefont {Zhao}\ \emph {et~al.}(2010)\citenamefont {Zhao},
  \citenamefont {Koschny},\ and\ \citenamefont {Soukoulis}}]{Zhao2010}%
  \BibitemOpen
  \bibfield  {author} {\bibinfo {author} {\bibfnamefont {R.}~\bibnamefont
  {Zhao}}, \bibinfo {author} {\bibfnamefont {T.}~\bibnamefont {Koschny}},\ and\
  \bibinfo {author} {\bibfnamefont {C.~M.}\ \bibnamefont {Soukoulis}},\
  }\href@noop {} {\bibfield  {journal} {\bibinfo  {journal} {Opt. Express}\
  }\textbf {\bibinfo {volume} {18}},\ \bibinfo {pages} {14553} (\bibinfo {year}
  {2010})}\BibitemShut {NoStop}%
\bibitem [{\citenamefont {Parappurath}\ \emph {et~al.}(2020)\citenamefont
  {Parappurath}, \citenamefont {Alpeggiani}, \citenamefont {Kuipers},\ and\
  \citenamefont {Verhagen}}]{Parappurath2020}%
  \BibitemOpen
  \bibfield  {author} {\bibinfo {author} {\bibfnamefont {N.}~\bibnamefont
  {Parappurath}}, \bibinfo {author} {\bibfnamefont {F.}~\bibnamefont
  {Alpeggiani}}, \bibinfo {author} {\bibfnamefont {L.}~\bibnamefont
  {Kuipers}},\ and\ \bibinfo {author} {\bibfnamefont {E.}~\bibnamefont
  {Verhagen}},\ }\href@noop {} {\bibfield  {journal} {\bibinfo  {journal} {Sci.
  Adv.}\ }\textbf {\bibinfo {volume} {6}},\ \bibinfo {pages} {eaaw4137}
  (\bibinfo {year} {2020})}\BibitemShut {NoStop}%
\bibitem [{\citenamefont {Huang}\ \emph {et~al.}(2019)\citenamefont {Huang},
  \citenamefont {Guo}, \citenamefont {Feng}, \citenamefont {Yu}, \citenamefont
  {Jiang}, \citenamefont {Zhang}, \citenamefont {Wang},\ and\ \citenamefont
  {Chen}}]{Huang2019}%
  \BibitemOpen
  \bibfield  {author} {\bibinfo {author} {\bibfnamefont {Q.}~\bibnamefont
  {Huang}}, \bibinfo {author} {\bibfnamefont {Z.}~\bibnamefont {Guo}}, \bibinfo
  {author} {\bibfnamefont {J.}~\bibnamefont {Feng}}, \bibinfo {author}
  {\bibfnamefont {C.}~\bibnamefont {Yu}}, \bibinfo {author} {\bibfnamefont
  {H.}~\bibnamefont {Jiang}}, \bibinfo {author} {\bibfnamefont
  {Z.}~\bibnamefont {Zhang}}, \bibinfo {author} {\bibfnamefont
  {Z.}~\bibnamefont {Wang}},\ and\ \bibinfo {author} {\bibfnamefont
  {H.}~\bibnamefont {Chen}},\ }\href@noop {} {\bibfield  {journal} {\bibinfo
  {journal} {Laser \& Photonics Reviews}\ }\textbf {\bibinfo {volume} {13}},\
  \bibinfo {pages} {1800339} (\bibinfo {year} {2019})}\BibitemShut {NoStop}%
\end{thebibliography}
\end{document}